\documentclass[pre,pr1,reprint,twocolumn,amsmath,10pt,amssymb,aps,superscriptaddress,showkeys,showpacs]{revtex4-1}
\usepackage[utf8]{inputenc}
\usepackage{lipsum}
\usepackage{enumitem,mathtools}
\usepackage{graphicx}% Include figure files
\usepackage{subcaption}
\usepackage{dcolumn}% Align table columns on decimal point
\usepackage{array}
\usepackage{bm}% bold math
\usepackage[colorlinks=true,
            linkcolor=blue,
            urlcolor  = blue,
            citecolor = blue,
            anchorcolor = blue]{hyperref}% add hypertext capabilities
\usepackage[%showframe,%Uncomment any one of the following lines to test 
scale=0.7, marginratio={1:1, 2:3}, ignoreall,% default settings
text={7in,10in},centering,
margin=.6in,
total={6.5in,8.75in}, top=1.2in, left=0.5in, includefoot,
height=10in,a4paper,hmargin={1.5cm,0.8in},
]{geometry}
\usepackage{fancyhdr}
\usepackage{mwe}

\newcommand*{\Scale}[2][4]{\scalebox{#1}{$#2$}}%
\usepackage[utf8]{inputenc}
\newcommand{\be}{\begin{equation}}
\newcommand{\ee}{\end{equation}}
\usepackage{etoolbox} % for \appto
\usepackage{xcolor}
\usepackage[capitalize]{cleveref}

\makeatletter
\appto{\appendix}{%
  \@ifstar{\def\thesection{\unskip}\def\theequation@prefix{A.}}%
          {\def\thesection{\Alph {section}}}%
}
\makeatother
\begin{document}
%\preprint{APS/123-QED}
%\title{Diffusion with stochastic resetting inside a trapping potential:\\
%A transition in optimal strategy of first passage}% Force line breaks with
\title{First passage of a particle in a potential under stochastic resetting: \\
a vanishing transition of optimal resetting rate.} 

\author{Saeed Ahmad}
\email{saeedmalik@iitb.ac.in}%
\author{Indrani Nayak}
\email{inayak21@phy.iitb.ac.in}
\author{Ajay Bansal} 
\author{Amitabha Nandi}
\author{Dibyendu Das}
\affiliation{Physics Department, Indian Institute of Technology Bombay, Mumbai 400076, India}%
\date{\today}% It is always \today, today,
\begin{abstract}
First passage in a stochastic process may be influenced by the presence of an external confining potential, as well as ``stochastic resetting'' in which the process is repeatedly reset back to its initial position. Here we study the interplay between these two strategies, for a diffusing particle in an one-dimensional trapping potential $V(x)$, being randomly reset at a constant rate $r$. Stochastic resetting has been of great interest as it is known to provide an `optimal rate' ($r_*$) at which the mean first passage time is a minimum. On the other hand an attractive potential also assists in first capture process. Interestingly, we find that for a sufficiently strong external potential, the advantageous optimal resetting rate vanishes (i.e. $r_*\to 0$). We derive a condition for this {\it optimal resetting rate vanishing transition}, which is continuous.  We study this problem for various functional forms of $V(x)$, some analytically, and the rest numerically. We find that the optimal rate $r_*$ vanishes with the deviation from critical strength of the potential as a power law with an exponent $\beta$ which appears to be universal. 

%We establish this by studying a diffusing particle in a generic potential $V(x) = k x^n$, being randomly reset at a constant rate. We present analytically exact results for few special values of $n$, in the absence and presence of delayed restart after reset.  
%
%Using a numerical scheme for obtaining the mean first passage time from the backward differential equations for any generic potential, we study other values of $n$ as well as another potential with a barrier, which are difficult to treat analytically.  We find the exponent $\beta$ is robust for all these cases.                 
%%%%%%%%%%%%%%%%%%%%%%%%%%%%%%%%%%%%%%%%%%%%%%%%%%%%%%%%%%

\begin{description}
\item[PACS number(s)]05.40.-a,02.50.-r,02.50.Ey
%\verb+05.40.-a,02.30.Em,02.50.-r,05.70.Fh+
\end{description}
\end{abstract}
\maketitle
\section{\label{sec1}Introduction}

Survival and first passage problems are of great interest in stochastic process literature \cite{satyaCurrSc1999Persist,redner2001guide,ralf2014first}. Such questions have been studied in theory of random walks \cite{Lomholt_Tal2008,Benichou_Loverdo_2011,Bray_Staya_2013}, polymer and interface kinetics \cite{krug_staya1997,preStaya_Das2005,prldas2008}, chemical reactions \cite{Reuveni_Shlomi_2014_MicMenten,Reuveni_Shlomi2015_MicMenten}, diffusion in quenched flow fields \cite{Staya_2003_QuenchF, DDas_2006_QuenchF}, algorithmic problems \cite{Montanari_2002Algo}, and biophysics \cite{Tchou2014first,Roldan_2016RNA,DDas_2016DNA}. In the course of stochastic evolution of single or multi-particle systems,  first passage is said to occur when an event of crucial interest happens for the first time.  The distribution of timescales of the first occurrence of the event, as well as various cumulants of the distribution are of interest \cite{gard_book}. The mean first passage time (MFPT) is typically infinite for simple diffusive problems in open geometry,  but  finite in case of closed geometries and in the presence of spatially attractive potentials.     
 
Recently Stochastic Resetting (SR) in stochastic processes has become a topic of active research  
%\cite{Evan_Staya_2011,LukasStaya2014Levy,Staya_Sajib_2015Morkov,Sokolov_PRL_2018}
\cite{Evan_Staya_2011,LukasStaya2014Levy,Apoorva_ShamikG_2016,ApalKundu_Evan2016reset_t,Shilkev_2017_ContiRand_R-t,Bhat_Rener_2016,Staya_Sanjib_2015,Staya_Sajib_2015Morkov,Evan_Staya_2011Position,APal_2015Potential,Christo_2015_Reset_Boun,RoldanShamik_2017,Shamik_Staya_2014FlucInter,Shamik_Apoorva_2016FInter,Sokolov_PRL_2018, Reuveni_2016First, APAl_Reuveni_2017FirstR,A_Pal_Reuveni_PRL_2019_branch_reset, Tal_Robin_S_Reuveni_nature_2018}.
%\cite{,,,,,,,,,,}.
 In SR problems, a stochastic process is repeatedly returned to its initial position after random time intervals. This ensures that the process does not drift off very far from the initial position, and as a result a (non-equilibrium) steady state is attained at large times. In addition to this, the original stochastic process  may attempt a first passage event. A natural question is whether SR assists or impedes the process of first capture. For simple diffusion with SR at a constant rate, Evans and Majumdar \cite{Evan_Staya_2011} showed that the MFPT becomes finite  -- thus SR assists in first capture. Moreover there is an optimal resetting rate (ORR) at which the MFPT is a minimum. 
Since then, this phenomena has been studied in a variety of different scenarios with different rules for resetting.  
The resetting time can be completely deterministic or taken from a power-law distribution~\cite{Bhat_Rener_2016, Apoorva_ShamikG_2016}. Similarly, the rate may have an explicit time \cite{ApalKundu_Evan2016reset_t, Shilkev_2017_ContiRand_R-t} or position \cite{Evan_Staya_2011Position} dependance. 
%on both the nature of the resetting as well as the medium itself and has been studied extensively \cite{Bhat_Rener_2016, Apoorva_ShamikG_2016, ApalKundu_Evan2016reset_t, Shilkev_2017_ContiRand_R-t, Evan_Staya_2011Position}. 
Optimality of such resetting processes have been studied for multiple walkers \cite{Bhat_Rener_2016,Staya_Sanjib_2015}, and fluctuating interfaces ~\cite{Shamik_Staya_2014FlucInter,Shamik_Apoorva_2016FInter}. 
 For any process with constant resetting rate, it has been shown  that at ORR,  variance of the first passage times equal to square of MFPT \cite{Reuveni_2016First}. Furthermore for more general SR time distributions, many universal identities and inequalities have been derived \cite{APAl_Reuveni_2017FirstR}.

We recall that for simple diffusion in one-dimension, SR at constant rate produces an {\it advantage}, and an ORR exists where MFPT is minimum. Can this advantage be nullified? A simple way to  do that is by introducing an additional attractive potential $V(x) = k x^n$, with $k>0$ and $n \in (0, \infty)$,  which drives the particle towards  the capture site at $x=0$. There exists earlier studies of SR in the presence of diverse external potentials \cite{APal_2015Potential,Christo_2015_Reset_Boun,RoldanShamik_2017}. 
%The resetting problem in presence of an external potential has been studied earlier  \cite{APal_2015Potential,Christo_2015_Reset_Boun,RoldanShamik_2017} but those focussed on . 
% However the main question was how the steady state distribution and therefore the MFPT is affected in the presence of an external potentia
In the absence of a potential (i.e. for $n = 0$), SR is advantageous. Similarly for any $n > 0$, if the strength of the 
potential $k \rightarrow 0$, it is as good as a flat potential -- hence SR helps towards first passage as in the $n=0$ case.  On the other hand 
if $k \rightarrow \infty$, the particle would be driven to the origin by an enormous advective force and first passage would happen instantly -- no amount of resetting or any other strategy can make the first passage time any lower.  
But  for any finite $k$, it 
remains an open question whether SR would still be a helpful strategy towards speedy first passage. In this work we explore 
how the potential competes with SR for dominance and beyond a critical threshold strength, renders SR to be redundant. We find that by tuning and increasing the strength $k$, ORR can be made to vanish for $k$ greater than a threshold value $k_c(n)$. Thus in the presence of sufficiently strong attractive potential, SR does not help in first capture any more.
In this paper, we study this {\it ORR vanishing transition}, and find an universal behaviour in its vicinity --- ORR scales as $\sim (k_c(n) - k)^{\beta}$ for $k < k_c$, with $\beta = 1$ (independent of $n$). 
Following \cite{Reuveni_Shlomi2015_MicMenten}, we also study the transition with reset
followed by a stochastic time overhead (with mean time $\langle T_{\rm on}\rangle$), as would be expected in a Michaelis-Menten reaction scheme (MMRS).

%\red{Earlier studies of the classic resetting problem in the presence of an external potential  \cite{APal_2015Potential,Christo_2015_Reset_Boun,RoldanShamik_2017} primarily focussed on the steady state distributions and first passage. Similarly, the idea of ORR vanishing transition has also been explored  earlier \cite{Christo_2015_Reset_Boun,Reuveni_Shlomi2015_MicMenten}. However, unlike our paper, where we vary the strength of the potential, in \cite{Reuveni_Shlomi2015_MicMenten}, an ORR vanishing transition was discussed by varying $\langle T_{\rm on} \rangle$. The special case of $n=1$ has been independently studied in another recent work  \cite{Reuveni_Arxiv_2018}.}

The idea of ORR vanishing transition is not entirely new \cite{Christo_2015_Reset_Boun,Reuveni_Shlomi2015_MicMenten}. Unlike our paper, where we vary the strength of the potential, in \cite{Reuveni_Shlomi2015_MicMenten}, an ORR vanishing transition was discussed by varying $\langle T_{\rm on} \rangle$. The special case of $n=1$ has been independently studied in another recent work  \cite{Reuveni_Arxiv_2018}.

%,A_Pal_Reuveni_Arxiv_2018}}. 
%\textcolor{blue}{In the latter . 
   
In section II, we define the problem mathematically and derive the condition which solves for the transition point $k_c(n)$. Then we argue why an universal exponent $\beta = 1$ is expected. In section III, we demonstrate these facts through exact results for linear ($n=1$), harmonic ($n=2$) and box ($n \rightarrow \infty)$ potentials. In section IV, we extend some of the results to the case of reset
followed by refractory period. In section V, we present a numerical scheme to study the problem, and apply it to cases of the cubic ($n = 3$) and quartic ($n=4$) potentials, as well as a potential which is a non-monotonic function of $x$.  We provide concluding remarks in section VI.  

%%%%%%%%%%%%%%%%%%%%%%%%%%%%%%%%%%%%%%%%%%%%%%%%%%%%%%%%%%%%%%%%%%%%%%%%
%%%%%%%%%%%%%%%%%%%%%%%%%%%%%%%%%%%%%%%%%%%%%%%%%%%%%%%%%%%%%%%%%%%%%%%%
\section{\label{sec:sec1} The problem and some general results} 
We consider a diffusing particle (with diffusion constant $D$) initially at $x = x_0$, subject to an external attractive potential $V(x)=kx^{n}$ with $k > 0$ and $n \in (0,\infty)$.  There is an absorbing boundary at $x=0$.  Additionally the particle position is stochastically reset back to $x_0$ at a constant rate $r$.  We are interested in the first passage of the particle as a result of the interplay of SR and the potential.     
\par   
As is often done in  first passage problems \cite{gard_book}, one may start with the backward differential Chapman-Kolmogorov equation for the probability $Q(x,t)$ of the particle to survive till time $t$, starting from any initial position $x$:
\be
\frac{\partial Q}{\partial t}=
D\frac{\partial^{2} Q}{\partial x^{2}}- V^{\prime}(x) \frac{\partial Q}{\partial x}-rQ+rQ_{0},  
\label{eq:N_Pow_Eqn}
\ee 
where $Q \equiv Q(x,t)$, and $Q_0 \equiv Q(x_0,t)$. Note that for our problem, the spatial derivative of the potential $V^{\prime}(x)=k n x^{n-1}$, in the above equation. The initial condition is $Q(x,0)=1$, and the absorbing boundary condition is $Q(0,t)=0$. Note that $Q$ is finite, while its spatial derivatives vanish as $x\rightarrow \infty$.  On finding $Q(x,t)$, one may replace $x$ by $x_0$ (the particular specified initial position) and solve for $Q(x_0,t)$.  
\par
Taking Laplace transformation with respect to $t$, and defining $q\equiv q(x,s)=\int^{\infty}_{0}\mathrm{d}t Q(x,t)e^{-st}$, $q_0 \equiv q(x_0,s)$, 
and  $y(x,s) =\big[q(x,s) -\frac{rq_{0}+1}{(r+s)}\big]$, from Eq.~(\ref{eq:N_Pow_Eqn}) we get the following equation for the function $y$: 
\be
\frac{\mathrm{d}^{2} y}{\mathrm{d} x^{2}}-n\gamma  x^{n-1}\frac{\mathrm{d} y}{\mathrm{d} x}-\alpha^{2}y=0. 
\label{eq:N_Pow_Eqn2}
\ee
In the above equation, $\gamma =k/D$ and $\alpha=\sqrt{r+s/D}$.  The Eq.~(\ref{eq:N_Pow_Eqn2}) is not easy to solve for general $n$, except for some special cases.  One general observation can be made by converting the 
standard form of Eq.~(\ref{eq:N_Pow_Eqn2}) to its normal form, $u^{''} + f(x) u = 0$, where $y = u e^{\frac{\gamma }{2}x^{n}}$ and 
$f(x)=-\alpha^{2}-\frac{n^{2}\gamma^{2}x^{2(n-1)}}{4}\big[1-\frac{2(n-1)}{n\gamma x^{n}}\big]$. Since $f(x) < 0$ for increasing $x$, 
it follows from Sturm's theorem \cite{simmons2016differential} that $u$ has atmost one zero. From this it follows that $y$ is a monotonically increasing function of $x$, between $y(0,s)=-\frac{rq_{0}+1}{(r+s)}$ and $y(\infty,s)=0$ (which follow from initial conditions), without any zero crossing in between.   
\par
In any stochastic process with resetting, it has been shown quite generally \cite{Reuveni_2016First} that MFPT  
\be
\langle T_r \rangle = \frac{(1 - \tilde{F}(r))}{r \tilde{F}(r)},
\label{eq:mean_Tr}
\ee
where $\tilde{F}(s)$ is the Laplace transform of the first passage probability distribution in the corresponding problem `without resetting'.  For our problem $\tilde{F}(x,s) = 1 - s q_1(x,s)$, where $q_1(x,s) = {\rm{Lim}}_{r \rightarrow 0}~ q(x,s)$. Often $\langle T_r \rangle$ has a {\it minimum} at an ORR $r = r_*$, i.e. $r_* = \{ r | \langle T_{r_*} \rangle = {\rm min} \langle T_r \rangle  \}$.  In the current problem, tuning the strength $k$ of the potential, it may be made to dominate over SR and thus make ORR vanish, i.e. $r_* \rightarrow 0$.  Near the latter transition point, since $r_*$ would be small,  one may approximate MFPT in Eq.~(\ref{eq:mean_Tr}) as a series in $r$ up to $O(r^2)$ (\textcolor{blue}{see Appendix \ref{APP:1} for $O(r^3)$}): 
\be
\langle T_r \rangle = \langle T \rangle  - r \frac{(\sigma^2 - \langle T \rangle^2)}{2} + r^2 (\frac{1}{6} \langle T^3 \rangle -\sigma^2 \langle T \rangle). %+ O(r^3)
\label{eq:MFPT_taylor}
\ee
Here the various moments on the right of Eq.~(\ref{eq:MFPT_taylor}) are for first passage times `without resetting';  similar expansions have been studied earlier \cite{Reuveni_Shlomi2015_MicMenten,Reuveni_2016First}. We note that in the limit of small $r_*$, derivative of Eq.~(\ref{eq:MFPT_taylor}), i.e. $d \langle T_r \rangle/dr |_{r = r_*} = 0$ yields (\textcolor{blue}{see Appendix}~\ref{APP:1}):
\be
r_* = \frac{1}{4} \frac{(\sigma^2 - \langle T \rangle^2)}{(\frac{1}{6} \langle T^3 \rangle -\sigma^2 \langle T \rangle)}. 
\label{eq:r_*}
\ee
This would imply two things. Firstly,  the ORR vanishing transition ($r_* \rightarrow 0$), coincides with the condition 
\be
\sigma^2  = \langle T \rangle^2, 
\label{eq:ORR_condition}
\ee
that is when `without resetting', variance of first passage times due to the tuned potential attains the same value as the square of the MFPT.  This means that the potential strength $k = k_c(n)$ at which the transition happens, may be solved from Eq.~(\ref{eq:ORR_condition}). Secondly if $r_*$ vanishes continuously, the expression on the right of Eq~(\ref{eq:r_*}) is expected to scale as follows: 
\be
r_* \sim (k_c(n) - k)^{\beta}.  
\label{eq:beta}
\ee 
If an analytic Taylor expansion of $r_*$ exists in $(k_c(n) - k)$ with first term non-vanishing, we would expect the exponent $\beta = 1$.  We would see below that this appears to be  true for all the potentials we consider. 
In what follows, we would often use dimensionless counterparts of $k$ and $r_*$, namely   $K = \big(\frac{k}{D}\big)^{\frac{1}{n}}x_{0}$ and $z^{*2} = r_*x_0^2/D$. 

In addition to a stochastic process (which for our case is a random walk in a potential) and SR, in a chemical MMRS, there is typically
an inert period after reset, with a mean time $\langle T_{\rm on} \rangle$. The latter problem has been studied generally in \cite{Reuveni_Shlomi_2014_MicMenten,Reuveni_Shlomi2015_MicMenten,Reuveni_2016First}. Yet with the aim of deriving few exact results 
of our interest, we would note few relevant formulas from those works. The MFPT is:  
\be
\langle T_r \rangle = \frac{(r \langle T_{\rm on} \rangle + 1 - \tilde{F}(r))}{r \tilde{F}(r)},
\label{eq:mean_Tr_on}
\ee
and expanding this is small $r$ near the transition (like in  Eq.~(\ref{eq:MFPT_taylor})), we may set  $d \langle T_r \rangle/dr |_{r = r_*} = 0$, and obtain (for small $r_*$):
\be
r_* = \frac{1}{4} \frac{(\sigma^2 - \langle T \rangle^2 - 2 \langle T_{\rm on} \rangle \langle T \rangle)}{(\frac{1}{6} \langle T^3 \rangle -\sigma^2 \langle T \rangle - \frac{1}{2} \langle T_{\rm on} \rangle [\sigma^2  -  \langle T \rangle^2])}. 
\label{eq:r_*_on}
\ee

%\be
%r_* = \frac{1}{4} \frac{(\sigma^2 - \langle T \rangle^2 - 2 \langle T_{\rm on} \rangle \langle T \rangle)}{(\langle T\rangle^{3} + \frac{1}{6} \langle% T^3 \rangle - \langle T \rangle \langle T^2 \rangle - \frac{1}{2} \langle T_{\rm on} \rangle [\sigma^2  -  \langle T \rangle^2])}. 
%\label{eq:r_*_on}
%\ee
The condition to locate the ORR vanishing transition is then revised from Eq.~(\ref{eq:ORR_condition}) to the following:
 \be
\sigma^2  = \langle T \rangle^2 + 2 \langle T_{\rm on} \rangle \langle T \rangle.  
\label{eq:ORR_condition_on}
\ee
Again as noted in our discussion below Eq.~(\ref{eq:beta}), $r_* \sim (k_c - k)$. 

We would proceed below to study some special values of $n$ analytically exactly, and a couple of others numerically, to test our 
expectations in Eq.~(\ref{eq:ORR_condition}), Eq.~(\ref{eq:ORR_condition_on}), and Eq.~(\ref{eq:beta}).

%%%%%%%%%%%%%%%%%%%%%%%%%%%%%%%%%%%%%%%%%%%%%%%%%%%%%%%%%%%%%%%%%%%%%%%%%%%%%%%%%%%%%%
%%%%%%%%%%%%%%%%%%%%%%%%%%%%%%%%%%%%%%%%%%%%%%%%%%%%%%%%%%%%%%%%%%%%%%%%%%%%%%%%%%%%%%
%%%%%%%%%%%%%%%%%%%%%%%%%%%%%%%%%%%%%%%%%%%%%%%%%%%%%%%%%%%%%%%%%%%%%%%%%%%%%%%%%%%%%%
\section{\label{sec2}{Analytical results for ORR transition}}

\subsubsection{\label{Sub:section_n=1}\textbf{Linear Potential ($n=1$)}}

\begin{figure*}
    \centering
    \begin{subfigure}[b]{0.3\textwidth}
        \includegraphics[width =0.96\textwidth,height=0.18\textheight]{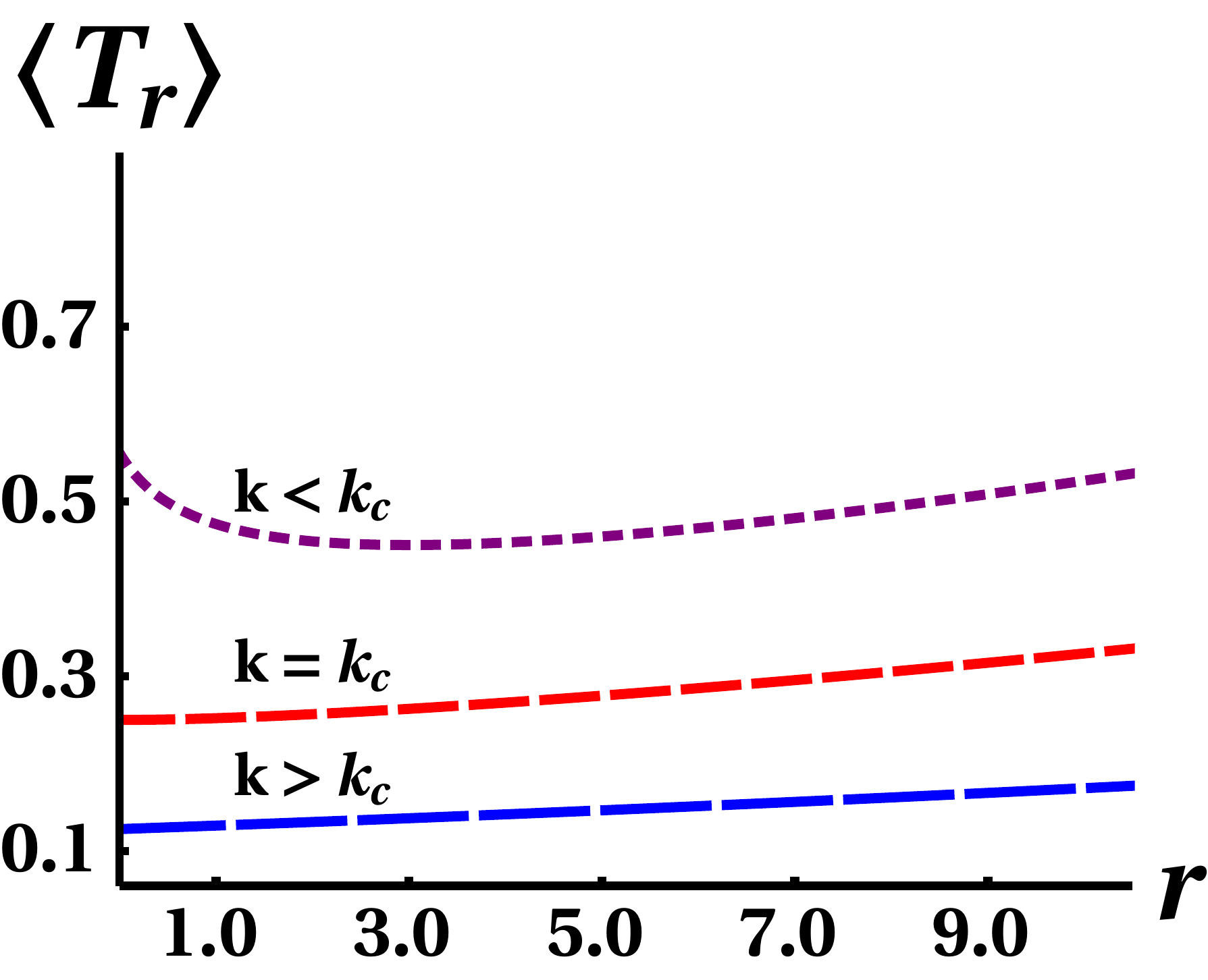}
        \caption{}
        %\caption{MFPT for linear potential.}
        \label{fig:Linear_P_MFPT}
    \end{subfigure}
    ~ %
    \begin{subfigure}[b]{0.3\textwidth}
        \includegraphics[width =0.96\textwidth,height=0.18\textheight]{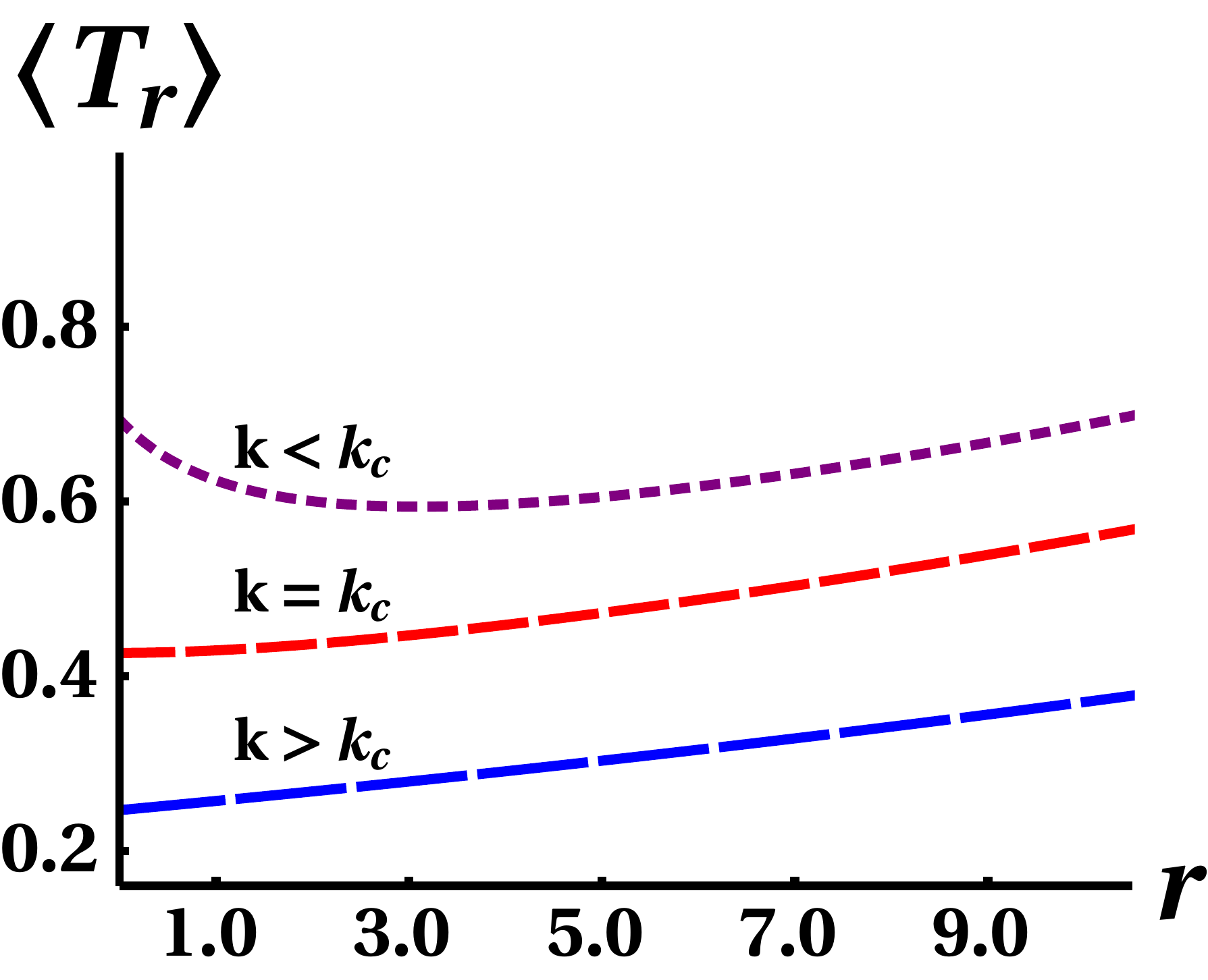}
        \caption{}
        \label{fig:Harmonic_P_MFPT}
    \end{subfigure}
    \begin{subfigure}[b]{0.3\textwidth}
        \includegraphics[width =0.96\textwidth,height=0.18\textheight]{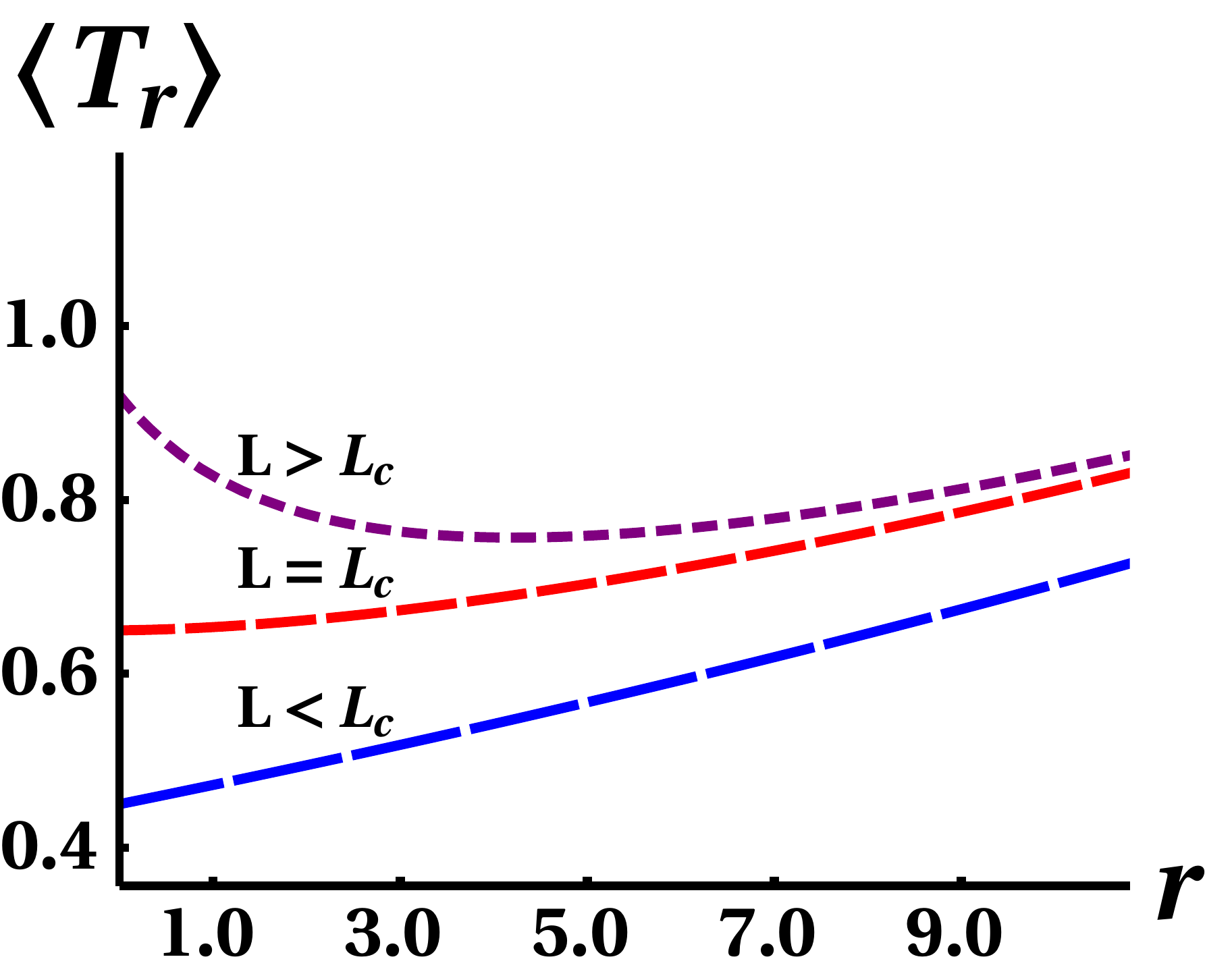}
        \caption{}
        \label{fig:Infinite_P_MFPT}
    \end{subfigure}
    \caption{We show the variation of MFPT with resetting rate $r$ for different $k$, in cases of (a) linear, (b) harmonic and (c) box potential. We used $x_{0}=0.5$, $D=0.5$ in the exact expressions for MFPT derived in the text. As may be seen, for $k \geq k_c$,  ORR $r_* = 0$.}
\label{fig:MFPT_T_r}
\end{figure*}

Substituting $n=1$ in Eq.~(\ref{eq:N_Pow_Eqn2}), we get the following:
\be
\frac{\mathrm{d}^{2} y}{\mathrm{d} x^{2}}-\gamma \frac{\mathrm{d} y}{\mathrm{d} x}-\alpha^{2}y=0,
\label{eq:diff_y_n=1}
\ee
the general solution for which is
\be
y(x,s)=A_1 e^{\big(\frac{\gamma }{2}+\sqrt{(\frac{\gamma }{2})^{2}+\alpha^{2}}\big)x}+B_1 e^{\big(\frac{\gamma }{2}-\sqrt{(\frac{\gamma }{2})^{2}+\alpha^{2}}\big)x}.
\label{eq:y_n=1}
\ee
The boundary conditions  $y(0,s)=-\frac{rq_{0}+1}{(r+s)}$ and $y(\infty,s)=0$ fix $A_1$ and $B_1$, and give solution for $y(x,s)$ and  hence $q(x,s)$ as follows: 
\be
q(x,s)=\bigg(\frac{rq_{0}+1}{r+s}\bigg)\bigg[1-e^{\big(\frac{\gamma }{2}-\sqrt{(\frac{\gamma }{2})^{2}+\alpha^{2}}\big)x}\bigg].
\label{eq:q_n=1}
\ee
Using Eq.~(\ref{eq:q_n=1}) or otherwise, without resetting (i.e. $r=0$),  $\tilde{F}(x_0,s) = \exp [\big(\frac{\gamma }{2}-\sqrt{(\frac{\gamma }{2})^{2}+\alpha_0^{2}}\big)x_0] $, where $\alpha_0 = \sqrt{\frac{s}{D}}$. 
This leads to $\langle T \rangle = -\frac{d{\tilde{F}}}{ds} |_{s=0} = \frac{x_0}{k}$, and
$\langle T^2 \rangle = \frac{d^2{\tilde{F}}}{ds^2} |_{s=0} = \langle T \rangle^2 + \frac{2D x_0}{k^3}$ . Then according to Eq.~(\ref{eq:ORR_condition}), the ORR vanishing transition happens at a threshold potential strength
\be
k_c = \frac{2D}{x_0}. 
\label{eq:kc_n=1}
\ee  
Note that arriving at the above result did not require us to refer to the actual problem with resetting. But we may also derive it by starting with the expression for MFPT under SR (i.e. $r > 0$):
\be
\langle T_r \rangle = q_0(x_0,s) |_{s=0} = \frac{1}{r}\bigg[e^{\big(\sqrt{(\frac{\gamma }{2})^{2}+\alpha^{2}_{0}}-\frac{\gamma }{2}\big)x_{0}}-1\bigg].
\label{eq:MFPT_n=1}
\ee
As may be seen from Fig.~(\ref{fig:Linear_P_MFPT}), the plot of $\langle T_r \rangle$ versus $r$ (following Eq.~(\ref{eq:MFPT_n=1})) has a minimum at $r = r_*(k)$ (ORR) for $k < k_c$, and for $k \geq k_c$ ORR is zero. 
The value of $r_*$ (for $k < k_c$) may be obtained from the  condition $d\langle T_r \rangle/dr |_{r=r_*}=0$ which is a transcendental equation as follows:
 \be
\frac{\frac{r_* x_0}{D}}{2\sqrt{({\frac{\gamma}{2})}^{2}+\frac{r_*}{D}}}=1-e^{(\frac{\gamma}{2} -\sqrt{({\frac{\gamma}{2}})^{2}+\frac{r_*}{D}})x_0}
\label{eq:ORR_n=1}
\ee
In Fig.~(\ref{fig:Linear_P_Z2_K}), a dimensionless ORR $z^{*2}$ is plotted against a dimensionless potential strength $K = \gamma x_0 = kx_0/D$, following Eq.~(\ref{eq:ORR_n=1}) (see the solid line). We see that the transition is at $K = K_c = 2$, i.e. $k_c = 2 D/x_0$, as we found in Eq.~(\ref{eq:kc_n=1}).  In the vicinity of $K = K_c$, for $K \leq K_c$, using the moments $\langle T \rangle$, $\langle T^2 \rangle$, and $\langle T^3 \rangle = -\frac{d^3{\tilde{F}}}{ds^3} |_{s=0} = \frac{12 D^2 x_0}{k^5} +\frac{6Dx^2_0}{k^4} + \frac{x_0^3}{k^3}$ in Eq.~\eqref{eq:r_*}, we get  $r_*$ and hence 
\be
z^{*2} = \frac{r_* x_0^2}{D} = \frac{3}{2}(K_{c}-K).
\label{eq:transition_n=1}  
\ee
The above Eq.~\eqref{eq:transition_n=1} may also be obtained from Eq.~(\ref{eq:ORR_n=1}) by expanding in small $r_*$ and $(K_c-K)$. 
This exact linearised form (plotted in Fig.~(\ref{fig:Linear_P_Z2_K}) as a dashed line) shows that the exponent $\beta = 1$, as expected in Eq.~(\ref{eq:beta}).

%%%%%%%%%%%%%%%%%%%%%%%%%%%%%%%%%%%%%%%%%%%%%%%%%%%%%%%%%%%%%%%%%%%%%%%%%%%%%%%%%%%%%%
%%%%%%%%%%%%%%%%%%%%%%%%%%%%%%%%%%%%%%%%%%%%%%%%%%%%%%%%%%%%%%%%%%%%%%%%%%%%%%%%%%%%%%
%%%%%%%%%%%%%%%%%%%%%%%%%%%%%%%%%%%%%%%%%%%%%%%%%%%%%%%%%%%%%%%%%%%%%%%%%%%%%%%%%%%%%%
%%%%%%%%%%%%%%%%%%%%%%%%%%%%%%%%%%%%%%%%%%%%%%%%%%%%%%%%%%%%%%%%%%%%%%%%%%%%%%%%%%%%%%

\subsubsection{\textbf{Quadratic potential ($n=2$)}}

For $n=2$, Eq.~(\ref{eq:N_Pow_Eqn2}) may be transformed by substituting $x=\sqrt{\xi/\gamma }$ and $y(x(\xi)) = w (\xi)$ to the familiar Confluent Hypergeometric equation \cite{simmons2016differential}:
\be
\xi\frac{\mathrm{d}^{2} w}{\mathrm{d} \xi^{2}}+(c -\xi) \frac{\mathrm{d} w}{\mathrm{d} \xi}- a w=0
\label{eq:diff_w_n=2}
\ee
with $c = 1/2$ and $a = \alpha^2/4\gamma$. 
The general solution in terms of Confluent Hypergeometric function $F_1(a,c; \xi)$ of first kind, is known. Transforming back to variable $x$, we have the general solution: 
\be
\Scale[0.95]{
y(x)=A_2\bigg[F_{1}\bigg(\frac{\alpha^2}{4\gamma},\frac{1}{2}; \gamma x^{2}\bigg) +C_2 x \sqrt{\gamma} F_{1}\bigg(\frac{\alpha^2}{4\gamma}+\frac{1}{2},\frac{3}{2};\gamma x^{2}\bigg)\bigg]}
\label{eq:y_n=2}
\ee
Using the boundary condition  $y(\infty, s)=0$ and the known asymptotic form  $\lim_{x\to \infty}F_{1}(a,c,x)=\frac{\Gamma (c)}{\Gamma (a)}e^{x}x^{a-c}$  \cite{luke1969special}, we get  $C_2=-2\frac{\Gamma\big(\frac{\alpha^2}{4\gamma}+\frac{1}{2}\big)}{\Gamma\big(\frac{\alpha^2}{4\gamma}\big)}$. The boundary condition $y(0,s)=-\frac{rq_{0}+1}{(r+s)}$  implies $A_2=-\frac{rq_{0}+1}{(r+s)}$. Putting these together, we have 
\be
q(x,s)=\bigg(\frac{rq_{0}+1}{r+s}\bigg)\bigg[1-G(x,s)\bigg]
\label{eq:q_n=2}
\ee
where 
\be 
\Scale[0.92]
{G =\bigg[F_{1}\bigg(\frac{\alpha^2}{4\gamma};\frac{1}{2};\frac{kx^{2}}{D}\bigg)-2\frac{\Gamma\big(\frac{\alpha^2}{4\gamma}+\frac{1}{2}\big)}{\Gamma\big(\frac{\alpha^2}{4\gamma}\big)} x \sqrt{\gamma} F_{1}\bigg(\frac{\alpha^2}{4\gamma}+\frac{1}{2};\frac{3}{2};\frac{kx^{2}}{D}\bigg)\bigg]}.
\label{eq:G_n=2}
\ee

\begin{figure}[hb!]
  \begin{minipage}{.5\textwidth}
    \begin{subfigure}[b]{0.45\textwidth}
      \includegraphics[width = 1.0\textwidth,height=0.13\textheight]{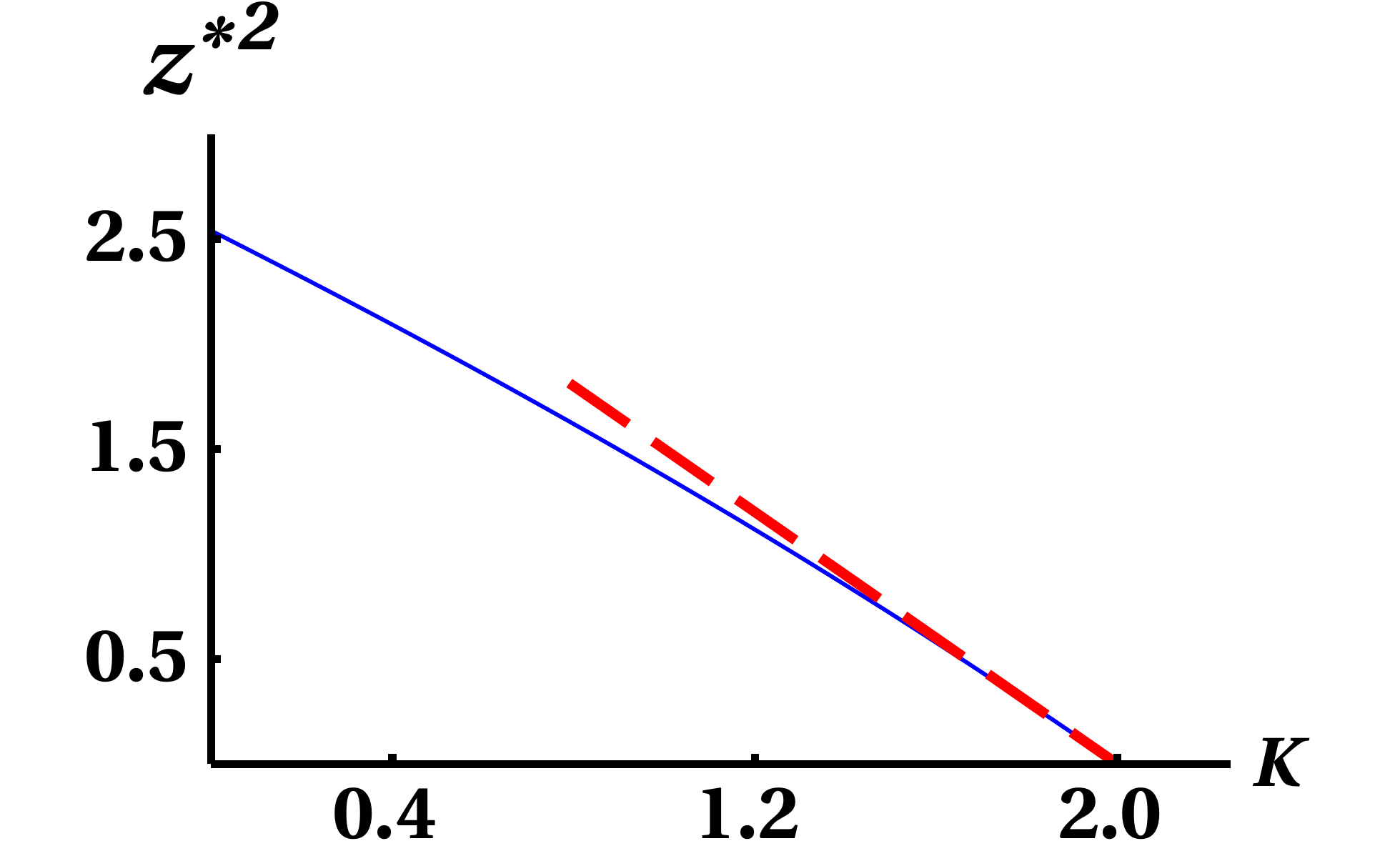}
      \caption{}\label{fig:Linear_P_Z2_K}
    \end{subfigure}
    \begin{subfigure}[b]{0.45\textwidth}
      \includegraphics[width = 1.0\textwidth,height=0.13\textheight]{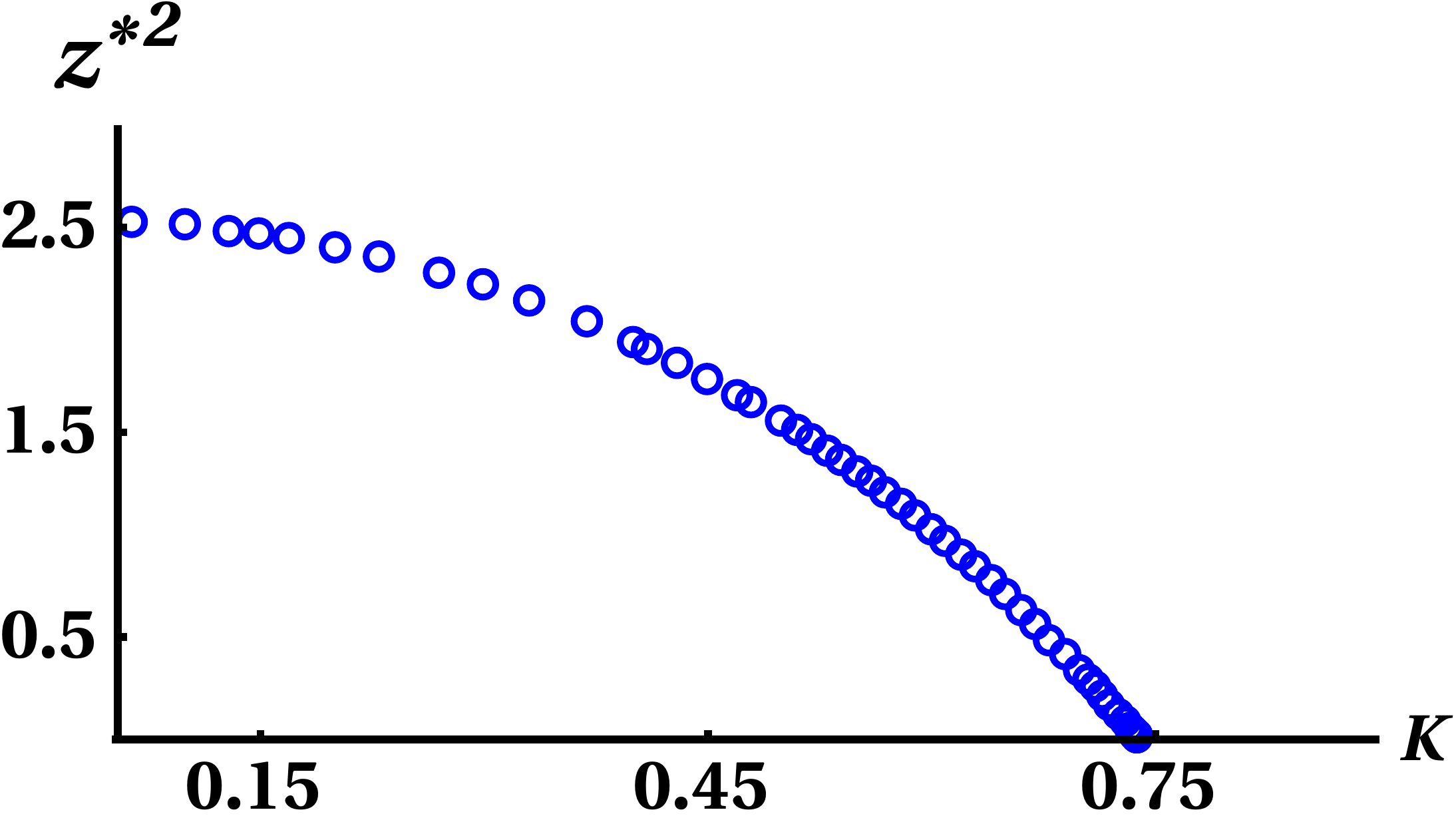}
      \caption{}\label{fig:Harmonic_P_Z2_K1}
    \end{subfigure}
    \begin{subfigure}[b]{0.45\textwidth}
        \includegraphics[width = 1.0\textwidth,height=0.13\textheight]{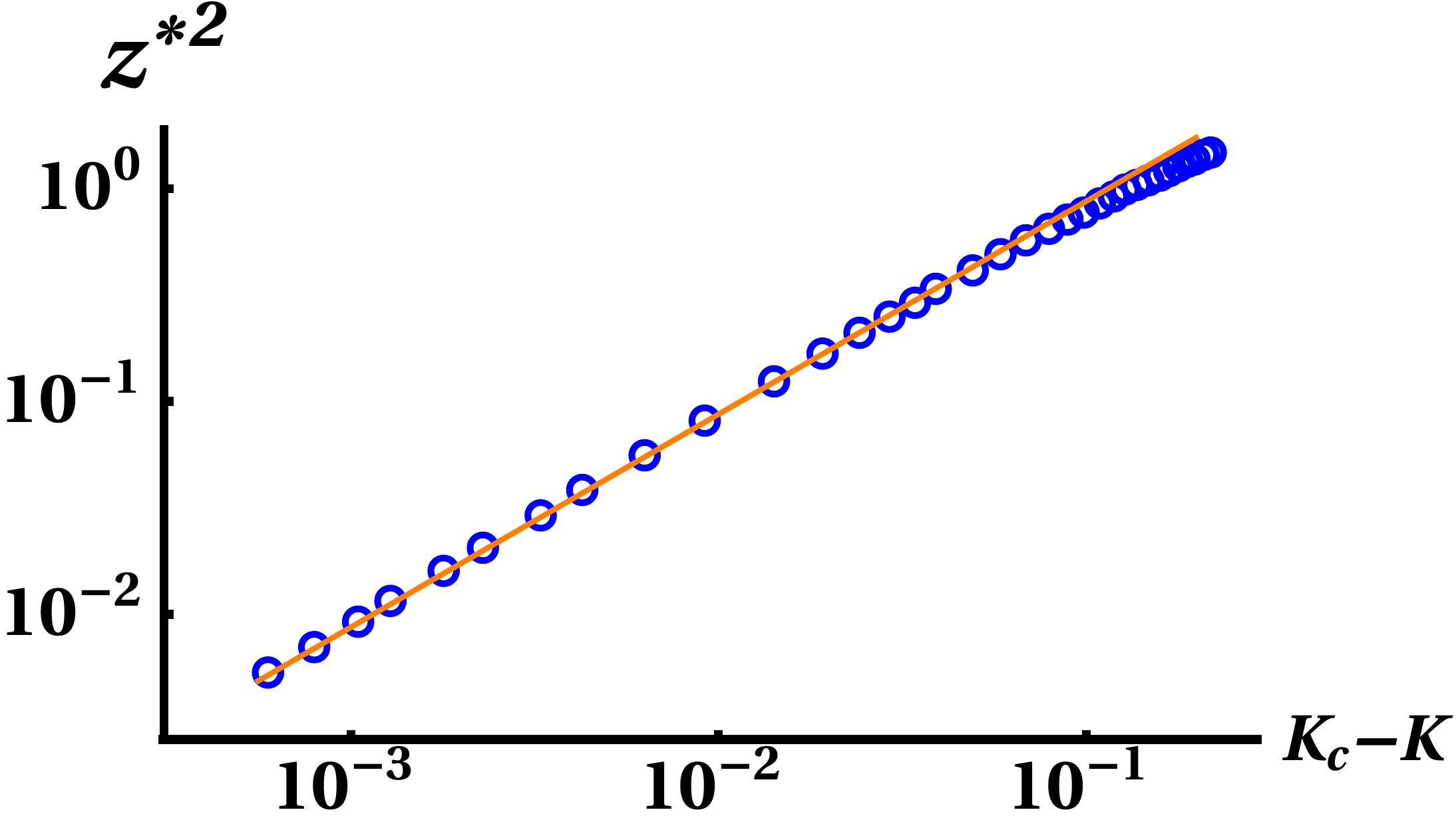}
        \caption{}
        \label{fig:Harmonic_P_Z2_K2}
    \end{subfigure}
    \begin{subfigure}[b]{0.45\textwidth}
      \includegraphics[width = 1.0\textwidth,height=0.13\textheight]{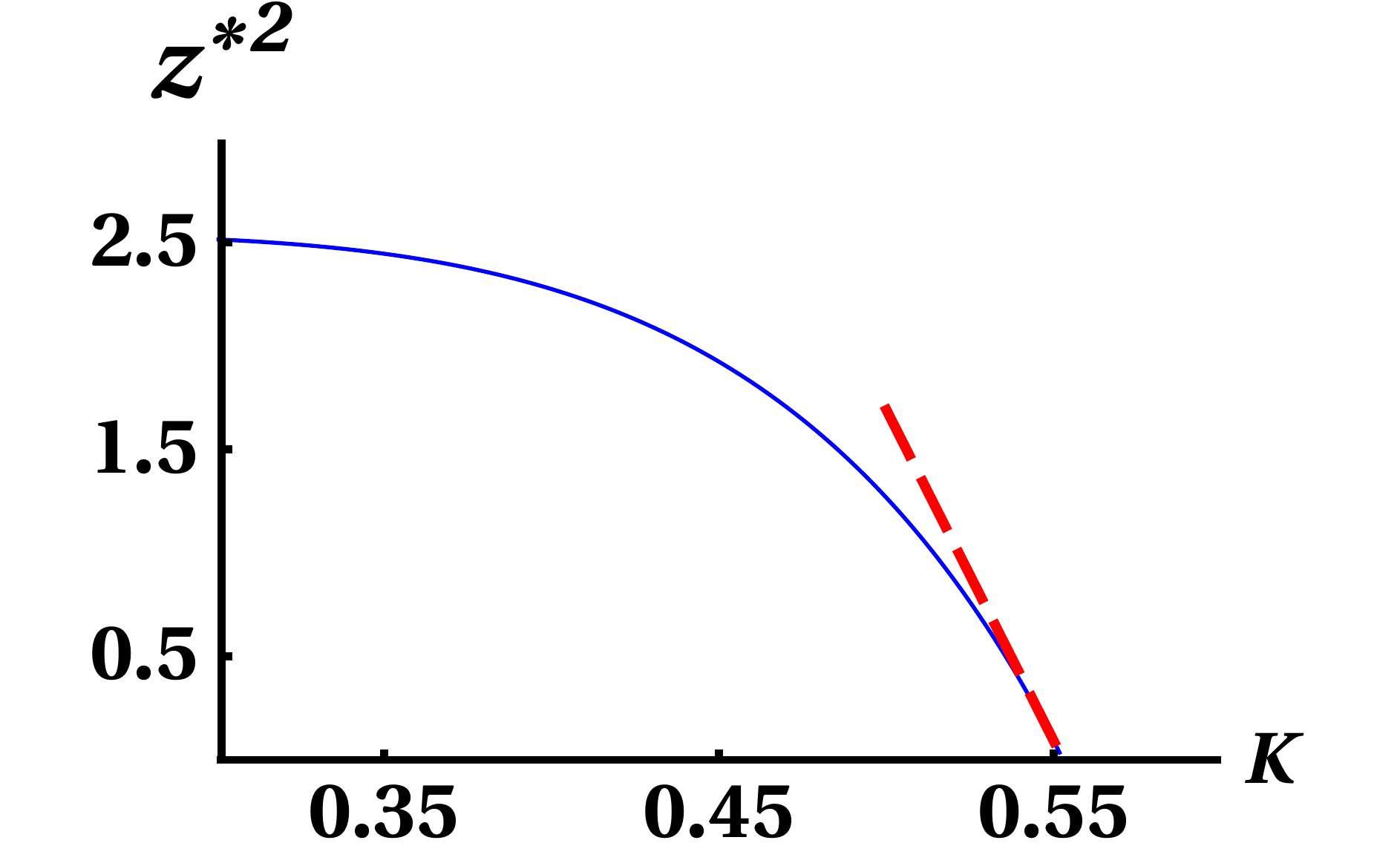}
      \caption{}\label{fig:Infinite_P_Z2_K}
    \end{subfigure}
    \caption{We show  $z^{*2 } = r_*x_0^2/D$ versus $K$  for (a) linear, (b) harmonic and (d) box potentials respectively. The $K_{c}$ values are 2, (0.7393$\pm$0.0001) and $1-1/\sqrt{5}$ corresponding to $n=1$, $2$ and $\infty$. For the harmonic case, 
    %since we do not have an analytic expansion  of $z^{*2}$ vs. small $(K_c - K)$, 
    we show in (c) the linearity in a log-log scale -- a solid line with power $1.0$ is put against the data.}
    \label{fig:Quartic_Pot_Transition1}
  \end{minipage}
  \hfill
  \begin{minipage}{0.4\textwidth}
  \end{minipage}
\end{figure}

Note that for the problem without resetting, $\tilde{F}(x,s) = G(x,s) |_{r=0}$. One may proceed to get $\langle T \rangle$ and $\langle T^2 \rangle$ from $\tilde{F}$, but since derivatives of Gamma functions and Confluent hypergeometric functions with respect to their indices would be involved, Eq.~(\ref{eq:ORR_condition}) for the location of the transition point $k_c$  is given by a somewhat cumbersome transcendental equation. Instead of treating that, we first derive the MFPT using Eq.~(\ref{eq:q_n=2}) and Eq.~(\ref{eq:G_n=2}) for $r > 0$ as follows:  
\be
\langle T_r \rangle= q_0(x_0,s) |_{s=0} = \frac{1}{r}\bigg[\frac{1}{G(x_{0},s=0)}-1\bigg].
\label{eq:MFPT_n=2}
\ee
In Fig.~(\ref{fig:Harmonic_P_MFPT}) we have plotted $\langle T_r \rangle$ against $r$ following Eq.~(\ref{eq:MFPT_n=2}), and we see that for $k < k_c$, there is a minimum at some $r = r_*(k) > 0$ (ORR). For $k \geq k_c$, ORR is zero. Beyond this we proceed numerically. We find the value of $r_*(k)$ within accuracy of $10^{-6}$, and plot its rescaled dimensionless counterpart $z^{*2}$ against dimensionless potential strength $K = x_0 \sqrt{k/D}$ in Fig.~(\ref{fig:Harmonic_P_Z2_K1}). This helps us locate $k_c$ (and $K_c \simeq 0.7393$) numerically. In the subfigure Fig.~(\ref{fig:Harmonic_P_Z2_K2}) we plot $z^{*2}$ versus $(K_c - K)$ in log-log scale for data values of $K$ very close to $K_c$. The expected power law with power $\beta = 1$ is shown.  
%%%%%%%%%%%%%%%%%%%%%%%%%%%%%%%%%%%%%%%%%%%%%%%%%%%%%%%%%%%%%%%%%%%%%%%%%%%%%%%%%%%%%%
%%%%%%%%%%%%%%%%%%%%%%%%%%%%%%%%%%%%%%%%%%%%%%%%%%%%%%%%%%%%%%%%%%%%%%%%%%%%%%%%%%%%%%
%%%%%%%%%%%%%%%%%%%%%%%%%%%%%%%%%%%%%%%%%%%%%%%%%%%%%%%%%%%%%%%%%%%%%%%%%%%%%%%%%%%%%%
%%%%%%%%%%%%%%%%%%%%%%%%%%%%%%%%%%%%%%%%%%%%%%%%%%%%%%%%%%%%%%%%%%%%%%%%%%%%%%%%%%%%%%

\subsubsection{\label{Sub:section_n=infty}\textbf{Box potential ($n=\infty$)}}

If we write $k = k_0/L^n$, then potential $V=k_0(\frac{x}{L})^{n}$. Now taking the limit $n \rightarrow \infty$, we have $V = 0$ for $x \leq L$ and $V = \infty$ for $x > L$, which is a box potential. In this limit, the modified form of the dimensionless potential strength is $\lim_{n\to\infty}K = \big(\frac{k_0}{DL^n}\big)^{\frac{1}{n}}x_{0} = \big(\frac{k_0}{D}\big)^{\frac{1}{n}}\frac{x_{0}}{L} \to \frac{x_0}{L}$. As $L$ becomes smaller, the strength $K$ rises, and the diffusing particle is more effectively confined and assisted towards the capture site $x=0$. The analog of  Eq.~(\ref{eq:N_Pow_Eqn2}) for this case is:
\be
\frac{\mathrm{d}^{2} y}{\mathrm{d} x^{2}} -\alpha^{2}y=0,
\label{eq:diff_y_n=infty}
\ee
with the boundary conditions, $y(0,s)=-\frac{rq_{0}+1}{(r+s)}$ and $\partial y/{\partial x} |_{x=L} = 0$. The general solution is $y = A_3 e^{\alpha x} + B_3 e^{-\alpha x}$, where $A_3$ and $B_3$ are fixed using the boundary conditions. This leads to:
\be
q(x,s)=\bigg(\frac{rq_{0}+1}{r+s}\bigg)\bigg[1-\frac{\cosh(\alpha(L-x))}{\cosh(\alpha L)}\bigg].
\label{eq:q_n=infty}
\ee

From Eq.~(\ref{eq:q_n=infty}) or otherwise, without SR, $\tilde{F}(x_0,s) = \cosh(\sqrt{\frac{s}{D}} (L-x_0))/\cosh(\sqrt{\frac{s}{D}} L)$. The latter implies that $\langle T \rangle = -\frac{d{\tilde F}}{ds} |_{s=0} =  \frac{x_0}{2D} (2L - x_0)$, and $\sigma^2 = \langle T^2 \rangle - \langle T \rangle^2 =\frac{ [L^4 - (L - x_0)^4]}{6 D^2}$. Substituting these in Eq.~(\ref{eq:ORR_condition}), we see that the transition value of the potential is given by 
\be
5 K_c^2 - 10 K_c + 4 = 0~~~~~~ \Rightarrow~ K_c = 1 - \frac{1}{\sqrt{5}}.
\label{eq:Kc_n=infty}
\ee
The other root in Eq.~(\ref{eq:Kc_n=infty}) is ignored as $K \leq 1$.  

The MFPT in this problem with SR is obtained from Eq.~(\ref{eq:q_n=infty}) as follows:
\be
\langle T_r \rangle = q_0(x_0,s) |_{s=0} = \frac{1}{r}\bigg[\frac{\cosh(\sqrt{\frac{r}{D}} L)}{\cosh(\sqrt{\frac{r}{D}} (L - x_{0}))}-1\bigg].
\label{eq:Tr_n=infty}
\ee
In Fig.~(\ref{fig:Infinite_P_MFPT}) we plot $\langle T_r \rangle$ against $r$ and see that the minimum at $r=r_*(k) > 0$ for $L > L_c$, 
vanishes for $L \leq L_c$. The exact expression for the $r_*$ (ORR for $K < K_c$) is given by $d \langle T_r \rangle/dr |_{r=r_*} = 0$ which lead to the following transcendental equation: 
\be 
\Scale[0.86]{
\begin{split}
\bigg(L \sinh\big(\sqrt{\frac{r_*}{D}} L\big)-\big(L - x_0\big)\tanh\big(\sqrt{\frac{r_*}{D}} (L - x_0)\big)\cosh\big(\sqrt{\frac{r_*}{D}} L\big)\bigg) \\
= \frac{2\sqrt{D}}{\sqrt{r_*}} \bigg(\cosh\big(\sqrt{\frac{r_*}{D}} L\big)-\cosh\big(\sqrt{\frac{r_*}{D}} (L - x_0) \big)\bigg).
\label{eq:ORR_n=infty}
\end{split}}
\ee
In Fig.~(\ref{fig:Infinite_P_Z2_K}) we plot $z^{*2}$ against $K$ (see solid line) following Eq.~(\ref{eq:ORR_n=infty}). The ORR vanishes at $K = K_c$ given by Eq.~(\ref{eq:Kc_n=infty}). In the limit of $K \to K_c$, Eq.\eqref{eq:r_*} yields $r_*$, using the moments $\langle T \rangle$, $\langle T^2 \rangle$, and $\langle T^3 \rangle = { \langle T \rangle [61L^4 -14 L^2 (L-x_0)^2 + (L-x_0)^4 ] }/{60D^2}$. This leads to 
\be
 z^{*2} = \frac{75(3\sqrt{5}-5)}{4}(K_{c}-K).
\label{eq:transition_n=infty}
\ee
The above exact linear form indicative of exponent $\beta = 1$, is shown as a dashed line in Fig.~(\ref{fig:Infinite_P_Z2_K}). 
%%%%%%%%%%%%%%%%%%%%%%%%%%%%%%%%%%%%%%%%%%%%%%%%%%%%%%%%%%%%%%%%%%%%%%%%%%%%%%%%%%%%%%
%%%%%%%%%%%%%%%%%%%%%%%%%%%%%%%%%%%%%%%%%%%%%%%%%%%%%%%%%%%%%%%%%%%%%%%%%%%%%%%%%%%%%%
%%%%%%%%%%%%%%%%%%%%%%%%%%%%%%%%%%%%%%%%%%%%%%%%%%%%%%%%%%%%%%%%%%%%%%%%%%%%%%%%%%%%%%
 
\section{Analytical results for ORR transition with stochastic time overhead}

In many stochastic processes with reseting, there may be a finite refractory period (with a mean $\langle T_{\rm on} \rangle$), as was discussed in the context of  MMRS  \cite{Reuveni_Shlomi2015_MicMenten}. In this section we discuss, how ORR vanishes on varying the potential strength $k$, for $\langle T_{\rm on} \rangle \neq 0$. The mean first passage time is given by Eq.\eqref{eq:mean_Tr_on}. The ORR is obtained by the condition $d\langle T_r \rangle/dr |_{r=r_*}=0$ which gives \cite{Reuveni_Shlomi2015_MicMenten}:
\be
r_*(1+r_*\langle T_{\rm on} \rangle)\frac{\partial \tilde{F}(r)}{\partial r}|_{r=r_*}= \tilde{F}(r_*)(\tilde{F}(r_*)-1)
\label{eq:r_*_Ton_ORR}
\ee
Furthermore, the ORR vanishing condition is given by Eq.~\eqref{eq:ORR_condition_on}.  Thus apart from expression of $r_*$ for $K < K_c$ (given by Eq.~\eqref{eq:r_*_Ton_ORR}), in the following we find the exact expressions of $K_c$ and small expansions of $r_*$ in terms of $(K_c - K)$ (using Eqs.~\eqref{eq:ORR_condition_on} and \eqref{eq:r_*_on}), for $n = 1$ and $n = \infty$. 

\begin{figure}
  \begin{minipage}{.5\textwidth}
    \begin{subfigure}[b]{0.45\textwidth}
      \includegraphics[width = 1.0\textwidth,height=0.10\textheight]{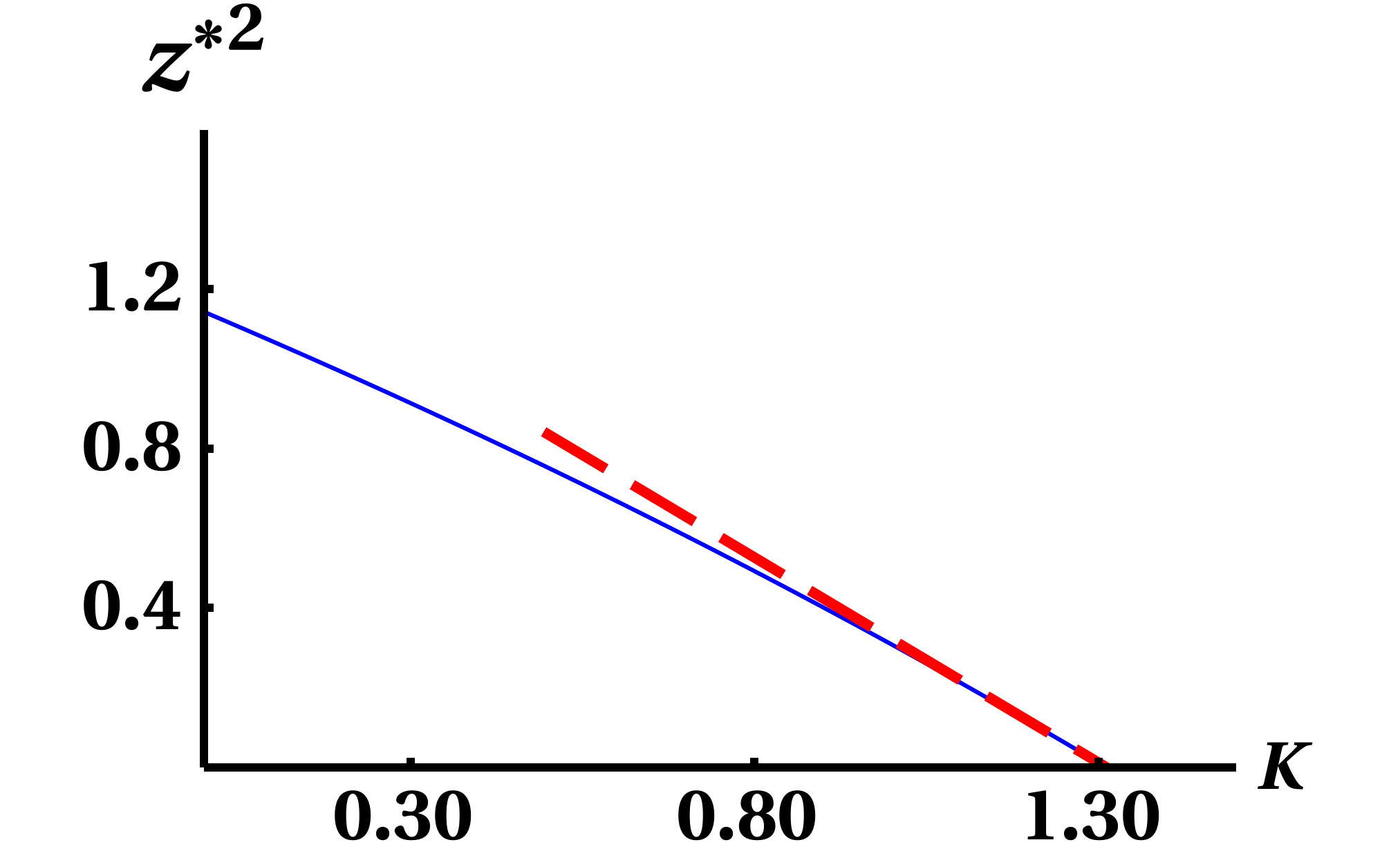}
      \caption{}\label{subfig:Linear_Pot_Z^2_K_Ton_0.1}
    \end{subfigure}
    \begin{subfigure}[b]{0.45\textwidth}
      \includegraphics[width = 1.0\textwidth,height=0.10\textheight]{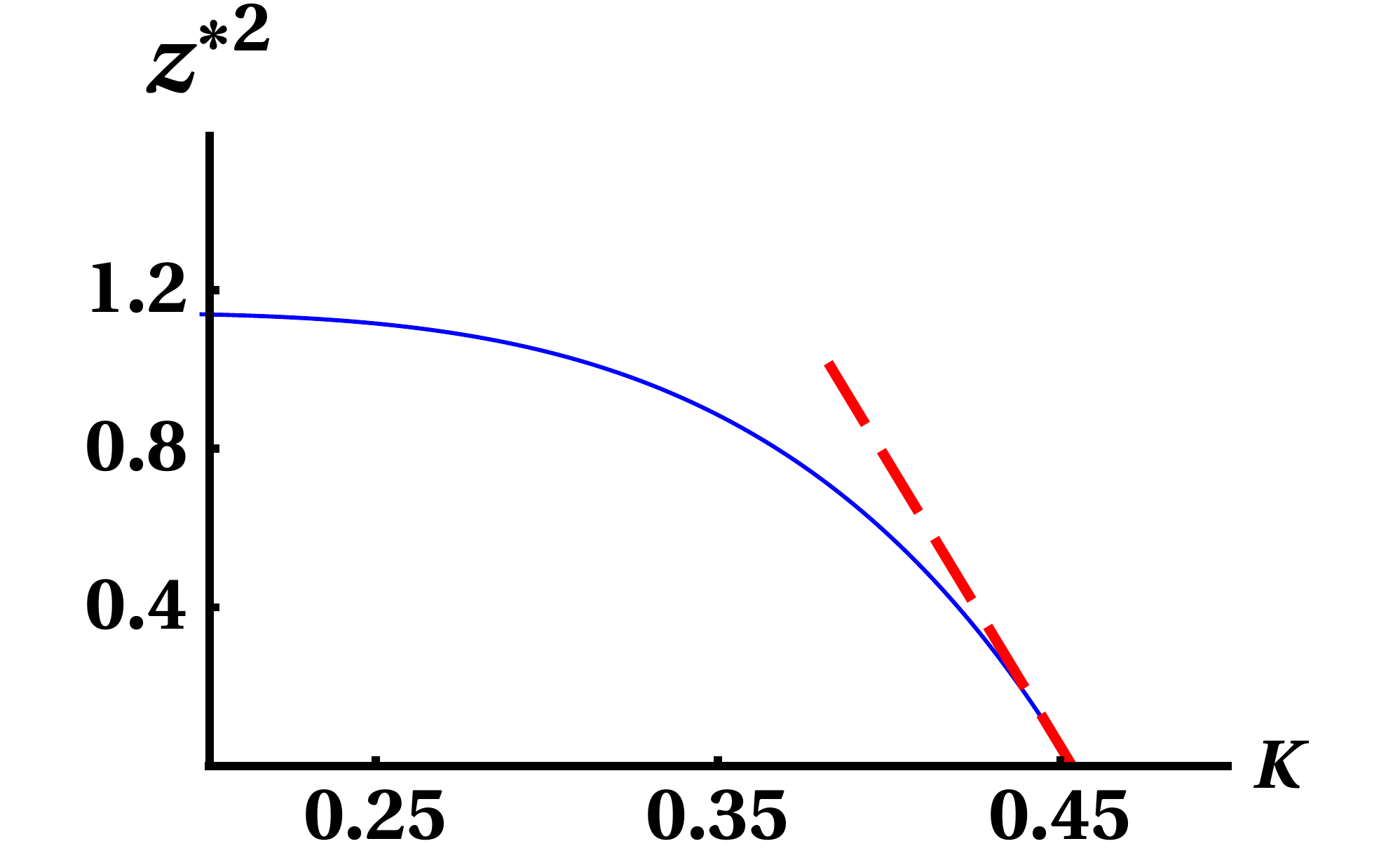}
      \caption{}\label{subfig:Infinite_Pot_Z^2_K_Ton_0.1}
    \end{subfigure}
    \caption{ We show $z^{*2 }$ vs. $K$ for linear potential and box potentials with $\langle T_{\rm on} \rangle=0.1$. The exact $K_{c}$ values are respectively, $1.3117$ and $0.4535$ obtained from equations in the text.}
    \label{fig:Linear_Infinite_Pot_Ton_0.1}
  \end{minipage}
  \hfill
  \begin{minipage}{0.4\textwidth}
  \end{minipage}
\end{figure}

For the linear potential ($n=1$), using $\tilde{F}(r) = e^{\big(\frac{\gamma }{2}-\sqrt{(\frac{\gamma }{2})^{2}+\frac{r}{D}}\big)x_{0}}$ from section \ref{Sub:section_n=1} in Eq.~\eqref{eq:r_*_Ton_ORR} we find $r_*$ and hence $z^{*2}$ as a function of $K$. A plot of this is shown in Fig.(\ref{subfig:Linear_Pot_Z^2_K_Ton_0.1}) (solid line) for  $\langle T_{\rm on} \rangle=0.1$. Then substituting the necessary moments (from section~\ref{Sub:section_n=1}) in Eq.~\eqref{eq:ORR_condition_on}, we find the exact transition point  
\be
K_c=\frac{4}{1+\sqrt{1+16D\langle T_{\rm on} \rangle/x_{0}^{2}}}, 
\label{eq:Kc_on_n=1}
\ee
which now depends on $\langle T_{\rm on} \rangle$ and is $< 2$ for any $\langle T_{\rm on} \rangle > 0$. Using the moments again in Eq.~\ref{eq:r_*_on}, we have  
\be
z^{*2} = \bigg[\frac{3(x^{4}_{0}+4Dx^{2}_{0}K_{c}\langle T_{\rm on} \rangle) ~~(K_{c}-K) }{2(x^{4}_{0}+6Dx^{2}_{0}K_{c}\langle T_{\rm on} \rangle+6D^{2}K^{2}_{c}{\langle T_{\rm on} \rangle}^{2})}\bigg],
\label{eq:ORR_n=1_Ton1} 
\ee
for small $r_*$ near $K_{c}$ indicating $\beta = 1$. 

Similarly for the box potential ($n=\infty$), using the function $\tilde{F}(r) = {\cosh(\sqrt{\frac{r}{D}} (L-x_0))}/{\cosh(\sqrt{\frac{r}{D}} L)}$ (from section \ref{Sub:section_n=infty}) in Eq.~\eqref{eq:r_*_Ton_ORR} we may obtain $z^{*2}$ versus $K$.  For $\langle T_{\rm on} \rangle=0.1$ 
a plot of this is shown in Fig.(\ref{subfig:Infinite_Pot_Z^2_K_Ton_0.1}) (solid line).  Again the relevant moments (from section~\ref{Sub:section_n=infty}) substituted in Eqs.(\ref{eq:ORR_condition_on}) gives, 
\be
K_c=\frac{4}{5+\sqrt{5+48D\langle T_{\rm on} \rangle/x_{0}^{2}}}, 
\label{eq:Kc_on_infty}
\ee
which is $ < (1 - 1/\sqrt{5})$ for any $\langle T_{\rm on} \rangle > 0$. Moreover as $K \to K_c$,  Eq.~\ref{eq:r_*_on} gives 
the linear form (with $\beta = 1$) for 
\be
\Scale[1.3]{
z^{*2}=\bigg[\frac{30x^4_0 K_c (4 - 5 K_c)~~(K_c-K)}{x^4_0 (11(1-K_c)^4 +6(1-K_c)^2-1)-72D^2{\langle T_{\rm on} \rangle}^2K^4_c} \bigg],
}
\label{eq:ORR_n=infty_Ton1}
\ee
which is shown as a dashed line in  Fig.(\ref{subfig:Infinite_Pot_Z^2_K_Ton_0.1}). 
%%%%%%%%%%%%%%%%%%%%%%%%%%%%%%%%%%%%%%%%%%%%%%%%%%%%%%%%%%%%%%%%%%%%%%%%%%%%%%%%%%%%%%
%%%%%%%%%%%%%%%%%%%%%%%%%%%%%%%%%%%%%%%%%%%%%%%%%%%%%%%%%%%%%%%%%%%%%%%%%%%%%%%%%%%%%%
%%%%%%%%%%%%%%%%%%%%%%%%%%%%%%%%%%%%%%%%%%%%%%%%%%%%%%%%%%%%%%%%%%%%%%%%%%%%%%%%%%%%%%
%%%%%%%%%%%%%%%%%%%%%%%%%%%%%%%%%%%%%%%%%%%%%%%%%%%%%%%%%%%%%%%%%%%%%%%%%%%%%%%%%%%%%%
%%%%%%%%%%%%%%%%%%%%%%%%%%%%%%%%%%%%%%%%%%%%%%%%%%%%%%%%%%%%%%%%%%%%%%%%%%%%%%%%%%%%%%

\section{\label{sec5}Numerical study of ORR transition in general potential $V(x)$}

Since analytical solutions are often difficult to find in case of arbitrary potentials $V(x)$, here we develop a numerical method to  
study the problem of ORR transition in such situations. The aim will be to obtain $\langle T_r \rangle$ numerically first as a function of $r$, and then locate its minimum ($r_*$) for a given potential strength.  Then one may vary the potential strength and study the corresponding variation and 
vanishing of $r_*$.  While $\langle T_r \rangle$ may be obtained using kinetic Monte-Carlo simulations~\cite{Kinetic_Monte_Carlo_1977}, that would typically have relatively high 
statistical fluctuations. Instead here we use a technique which is independent of statistical fluctuations.  

We note that $\langle T_r \rangle =  q(x_0,s)|_{s=0}$.  In section~\ref{sec:sec1}, we discussed that the Laplace transform of the survival probability $q(x_0,s)$ is related to another function $y(x_0,s) = q(x_0,s) + y(0,s)$, where  $y(0,s)= - \frac{r q(x_0,s) +1}{(r+s)}$. One knows the 
differential equation satisfied by $y(x,s)$ (namely Eq.~(\ref{eq:N_Pow_Eqn2}) for $V(x) = k x^n$) but its numerical solution is not straight forward, since its boundary condition $y(0,s)$ actually depends on the unknown $q(x_0,s)$ which we seek to find. This problem is avoided by studying instead a scaled function, $\tilde{y}(x,s)=y(x,s)/y(0,s)$, which has simpler boundary conditions $\tilde{y}(0,s)=1$ and $\tilde{y}(\infty,s)=0$ and satisfies the following equation:
\be
\frac{\mathrm{d}^{2} \tilde{y}}{\mathrm{d} x^{2}}=\frac{V^{'}(x)}{D}\frac{\mathrm{d} \tilde{y}}{\mathrm{d} x}+\frac{(r + s)}{D} \tilde{y}
\label{eq:NUmerical_Sol}
\ee
We solve this differential Eq.~(\ref{eq:NUmerical_Sol})  using {\it NDSolve} technique in {\it Mathematica} which includes {\it ExplicitRungeKutta} method to obtain $\tilde{y}(x,s)$. 
Since $q(x_0,s) = (\tilde{y}(x_0,s) - 1) y(0,s)$, we obtain 
\be
q(x_0,s) = \frac{1-\tilde{y}(x_0, s)}{s+r\tilde{y}(x_0,s)}. 
\ee  
Thus the knowledge of the numerically determined $\tilde{y}(x_0, s)|_{s=0}$ finally gives us  the desired mean first passage time 
\be
\langle T_r \rangle = q(x_0,s)|_{s=0} = \frac{1-\tilde{y}(x_0, 0)}{r\tilde{y}(x_0,0)}. 
\label{eq:Tr_n=general}
\ee
Sturm's theorem discussed in section~\ref{sec:sec1} ensures that $y(x_0,0)$ (and hence $\tilde{y}(x_0,0)$)
is non-zero for finite $x_0$, and Eq.\eqref{eq:Tr_n=general} is therefore well defined. We checked the reliability of this technique  by matching the exactly known $\langle T_r \rangle$ (for Linear and Harmonic potentials from Eqs.~\eqref{eq:MFPT_n=1} and \eqref{eq:MFPT_n=2}) to the numerically obtained $\langle T_r \rangle$ up to accuracy of $10^{-8}$. This precision 
of $\langle T_r \rangle$ corresponded to our choice of discrete step-size of $10^{-4}$ for variation of resetting rate $r$.  Thus all our answers 
below for values of $r_*$ are limited by this precision. We apply the numerical method below to study few different potentials. 
%%%%%%%%%%%%%%%%%%%%%%%%%%%%%%%%%%%%%%%%%
%%%%%%%%%%%%%%%%%%%%%%%%%%%%%%%%%%%%%%%%%
%%%%%%%%%%%%%%%%%%%%%%%%%%%%%%%%%%%%%%%%%
%%%%%%%%%%%%%%%%%%%%%%%%%%%%%%%%%%%%%%%%%
%%%%%%%%%%%%%%%%%%%%%%%%%%%%%%%%%%%%%%%%%

\begin{figure}
  \begin{minipage}{.5\textwidth}
    \begin{subfigure}[b]{0.45\textwidth}
      \includegraphics[width = 1.0\textwidth,height=0.15\textheight]{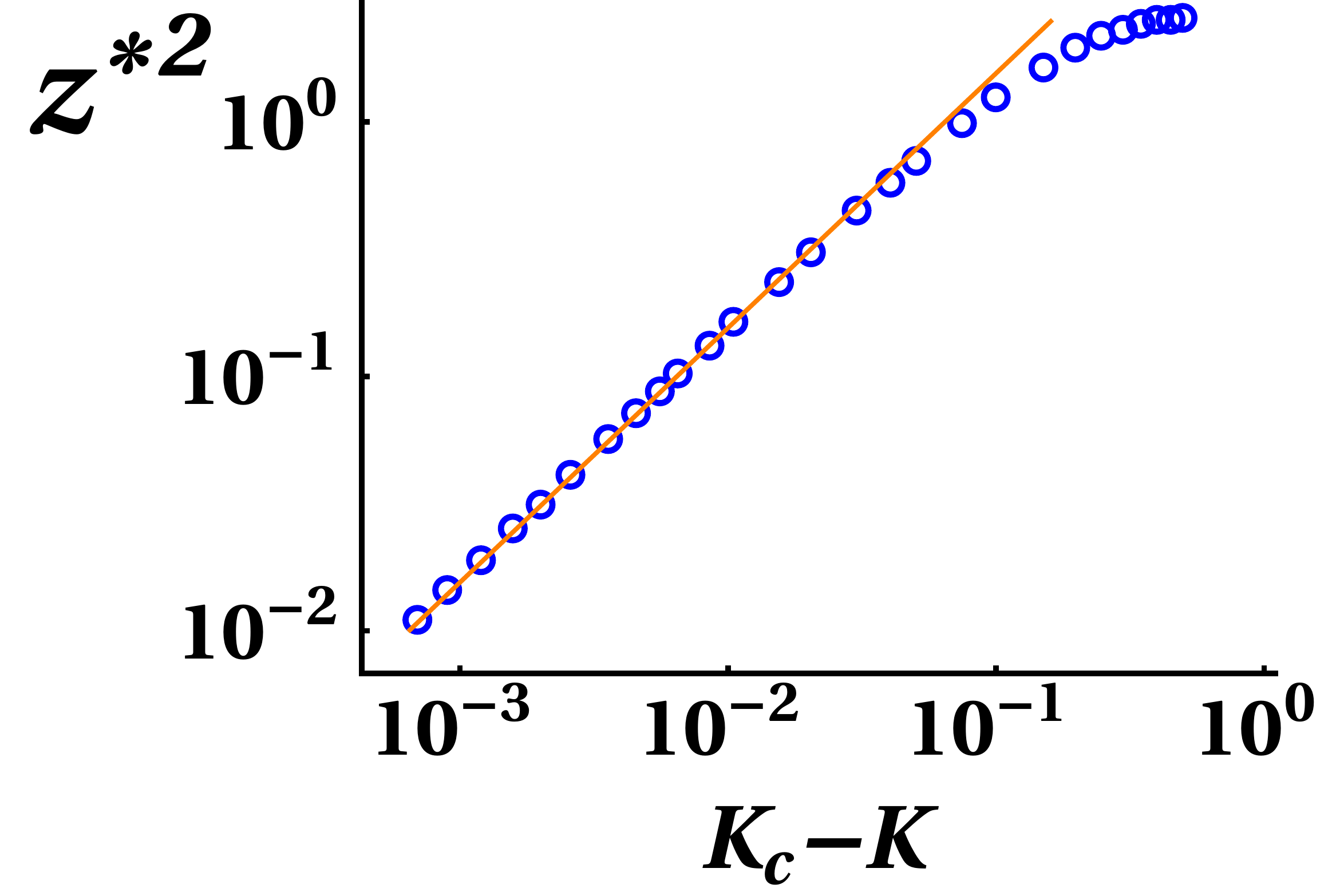}
      \caption{}\label{subfig:Cubic_Pot_Z^2_K}
    \end{subfigure}
    \begin{subfigure}[b]{0.45\textwidth}
      \includegraphics[width = 1.0\textwidth,height=0.15\textheight]{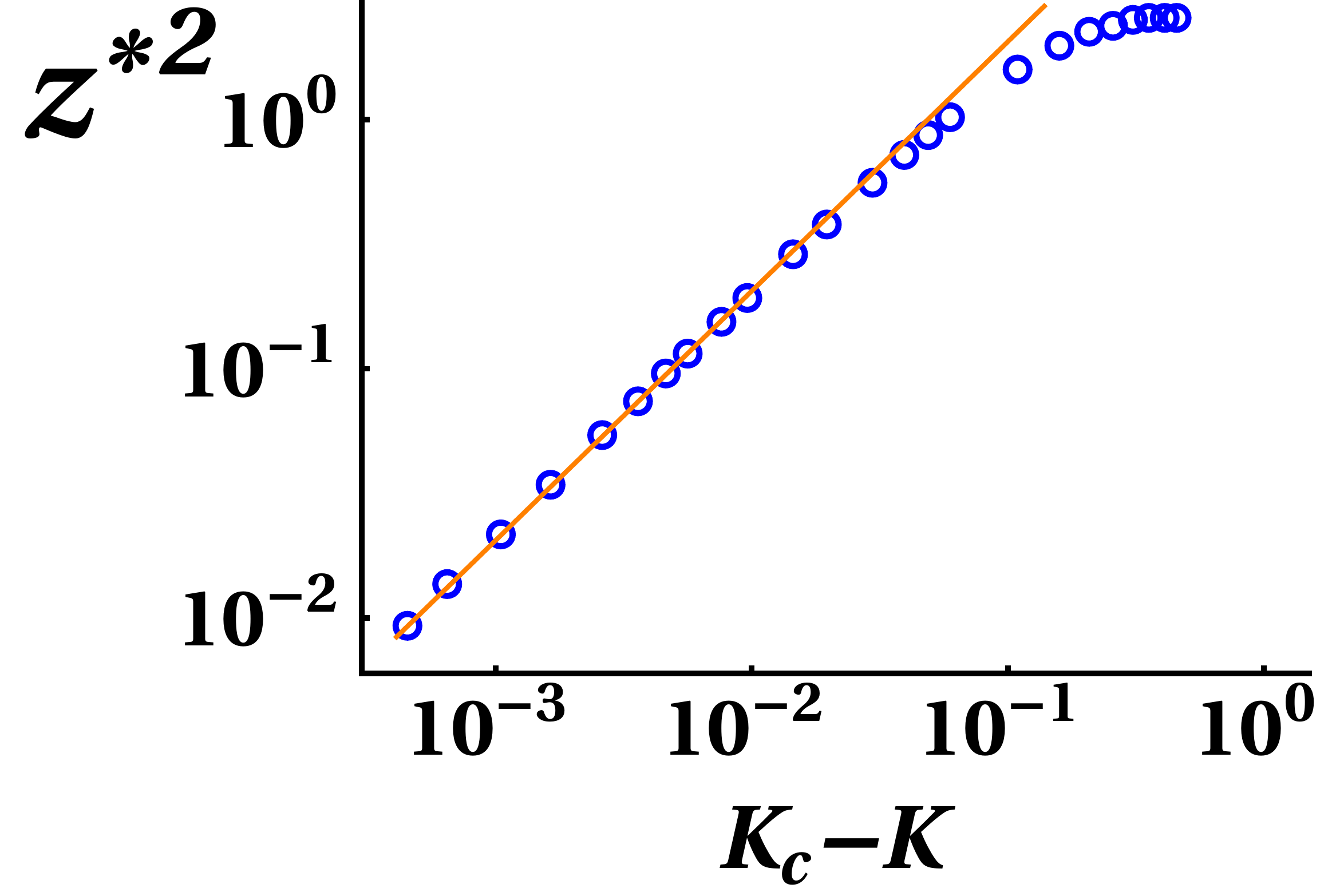}
      \caption{}\label{subfig:Quartic_Pot_Z^2_K}
    \end{subfigure}
    \caption{We show $z^{*2 }$ vs. $K$ for cubic power and quadratic potentials. The numerical $K_{c}$ values are respectively $0.6006 \pm 0.0001$ and $0.5597\pm 0.0001$.}
    \label{fig:Cubic_Quartic_Pot}
  \end{minipage}
  \hfill
  \begin{minipage}{0.4\textwidth}
  \end{minipage}
\end{figure}

{\it Cubic ($kx^{3}$) and Quartic ($kx^4$) potentials}:  For a chosen initial point $x_{0}=0.5$ and diffusion constant $D=0.5$, we obtain 
$\langle T_r \rangle$ as a function of $r$, and the corresponding optimal point $r_*$ for a given $k$.  Note that although $r_*$ and the 
value of $k = k_c$ where $r_*$ vanishes depends on $x_0$ and $D$, we express our results in terms dimensionless quantities 
$z^{*2}$ and $K$ which are supposed to be universal.  For the cubic case, with $K=\big(\frac{k}{D}\big)^{\frac{1}{3}}x_{0}$ we find that  
$K_c = 0.6006 \pm 0.0001$. In  Fig.~(\ref{subfig:Cubic_Pot_Z^2_K}), we plot $z^{*2}$ against $(K_c-K)$ in log-log plot and find 
a linear form valid over a couple of decades. For the quartic potential, with $K=\big(\frac{k}{D}\big)^{\frac{1}{4}}x_{0}$ we find that  
$K_c = 0.5597 \pm 0.0001$. In Fig.~(\ref{subfig:Quartic_Pot_Z^2_K}), we plot $z^{*2}$ against $(K_c-K)$ in log-log plot and find 
a linear form confirming again that $\beta = 1$.  

All the analytical and numerical results for various powers $n$ of the power law potential $V(x)$ may be summarised in Table (\ref{tab:Kc_behaviour}). We see that $K_c(n)$ decreases with $n$ and saturates as $n \to \infty$.
%%%%%%%%%%%%%%%%%%%%%%%%%%%%%%%%%%%%%%%%%%%%%%%%%%%%%%%%%%%%%%%%%%%%%%%%%%%%%%%%%%%%%%
%%%%%%%%%%%%%%%%%%%%%%%%%%%%%%%%%%%%%%%%%%%%%%%%%%%%%%%%%%%%%%%%%%%%%%%%%%%%%%%%%%%%%%
\begin{table}[ht!]
\centering
\begin{tabular}{ |p{2cm}|p{4cm}| } 
 \hline 
 Power $n$ &  $K_c = \big(\frac{k_c}{D}\big)^{\frac{1}{n}}x_{0}$  \\ 
\hline
  1 &  2 \\ 
\hline
  2 &  0.7393$\pm$0.0001  \\ 
\hline
  3 &  0.6006$\pm$0.0001  \\ 
 \hline
  4 &  0.5597$\pm$0.0001 \\
\hline
 $\infty$ & $\frac{x_0}{L_c} = 1- \frac{1}{\sqrt{5}} \simeq 0.5528 $ \\
\hline
\end{tabular}
\caption{The table contains $K_c$ with power $n$. }
\label{tab:Kc_behaviour}
\end{table}
\\
%%%%%%%%%%%%%%%%%%%%%%%%%%%%%%%%%%%%%%%%%%%%%%%%%%%%%%%%%%%%%%%%%%%%%%%%%%%%%%%%%%%%%%
%%%%%%%%%%%%%%%%%%%%%%%%%%%%%%%%%%%%%%%%%%%%%%%%%%%%%%%%%%%%%%%%%%%%%%%%%%%%%%%%%%%%%%

\begin{figure}[ht!]
  \begin{minipage}{.5\textwidth}
    \begin{subfigure}[b]{0.45\textwidth}
      \includegraphics[width = 1.0\textwidth,height=0.13\textheight]{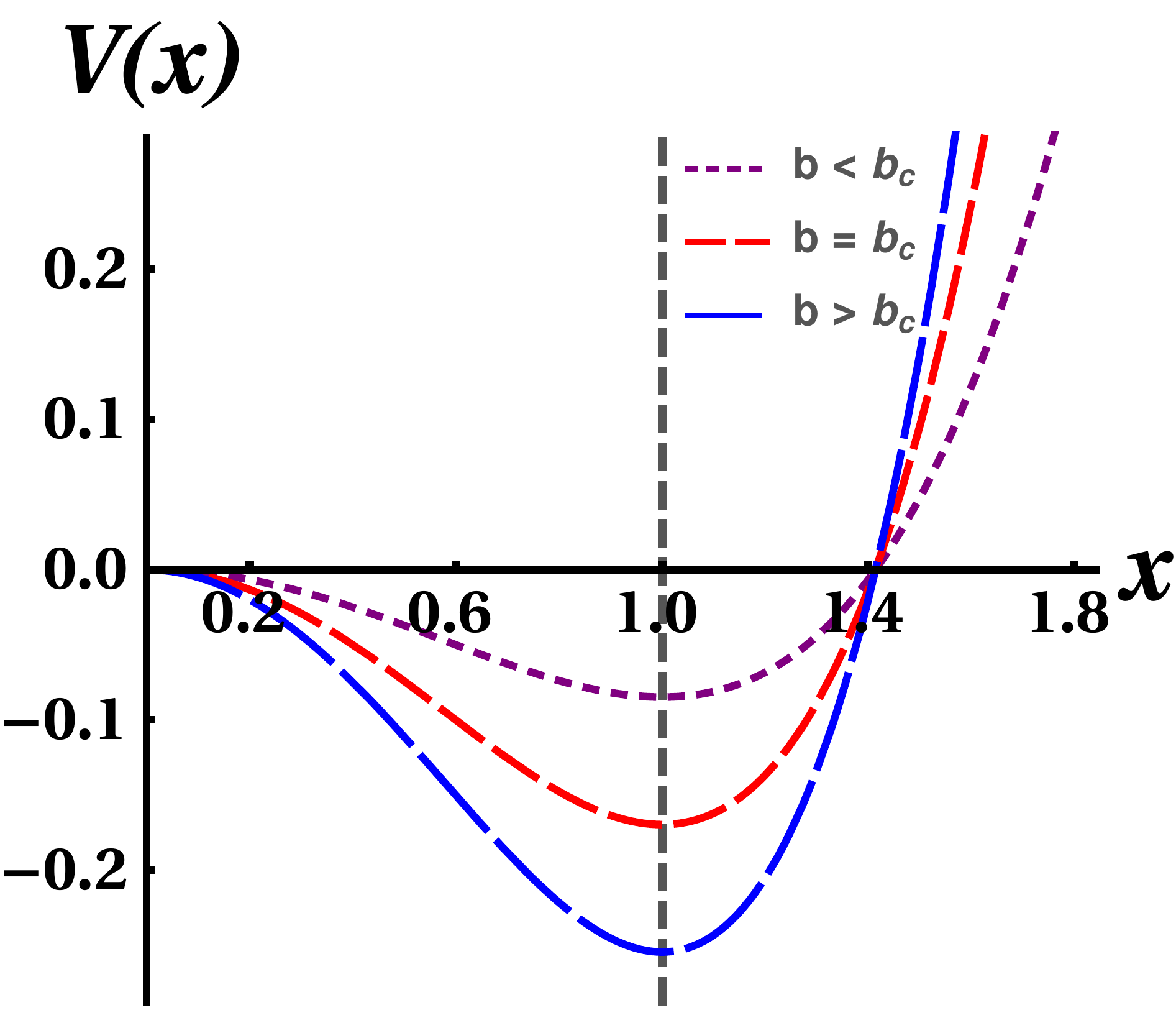}
      \caption{}\label{subfig:Mixed_Quartic_Pot}
    \end{subfigure}
    \begin{subfigure}[b]{0.45\textwidth}
      \includegraphics[width = 1.0\textwidth,height=0.13\textheight]{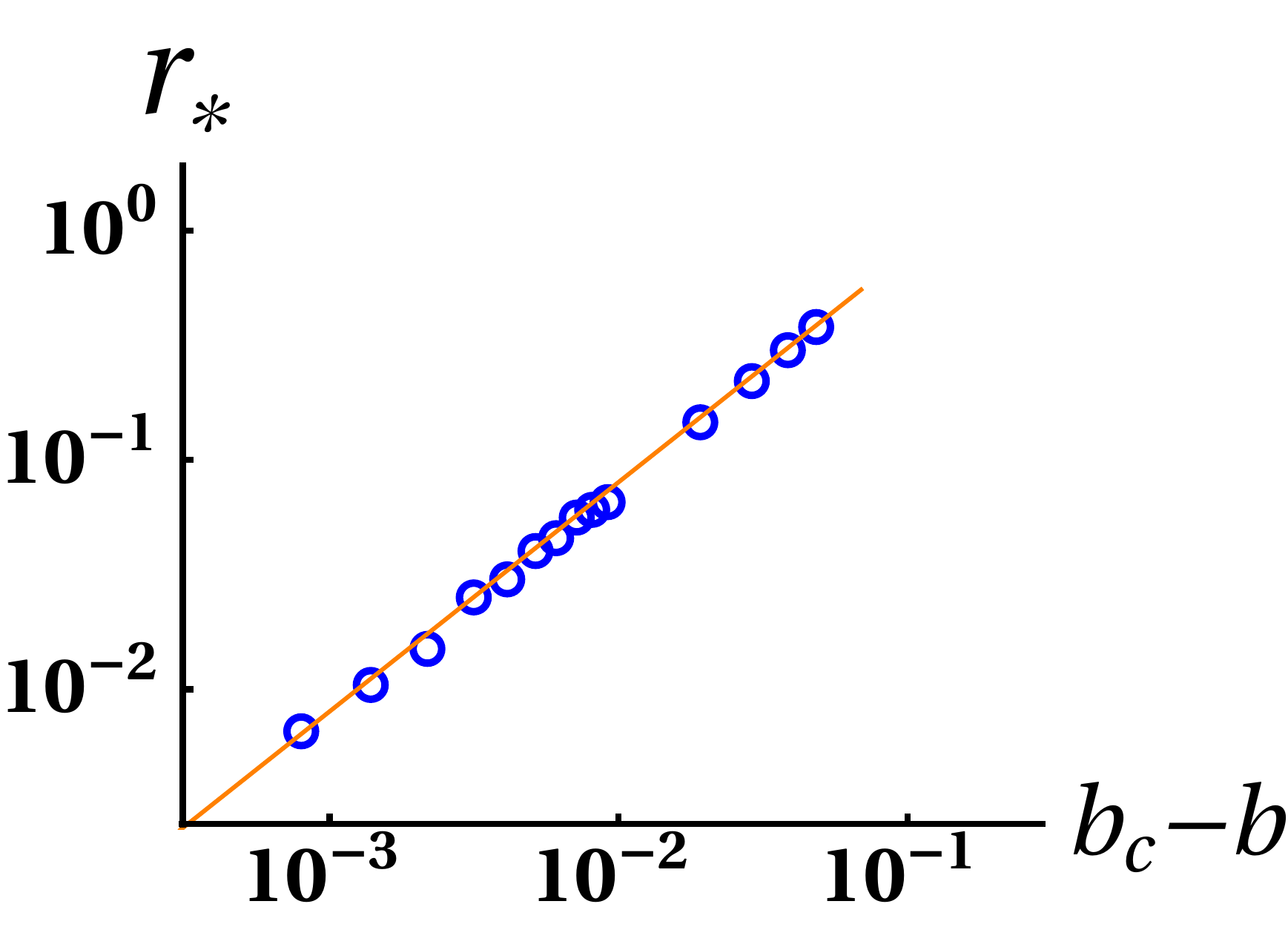}
      \caption{}\label{subfig:Mixed_Pot_Tran}
    \end{subfigure}
    \caption{We show in (a) the variation $V(x)$ with $x$ for various values of b.  The curves are for $b = 0.1696$, $b_c=0.3392$, and $b = 0.5088$. In (b) a log-log graph is shown for $r_*$ vs. $b_c-b$. Here $D=1$ and $x_0=1$. A line with power $1$ is in good agreement with the data.}
    \label{fig:Quartic_Pot_Transition}
  \end{minipage}
  \hfill
  \begin{minipage}{0.4\textwidth}
  \end{minipage}
\end{figure}

{\it A non-monotonic quartic potential}: In this part we study $\langle T_r \rangle$ vs $r$ for a one-dimensional potential whose behavior is somewhat different from the ones studied so far. The potential $V(x)=-\frac{b}{2}+\frac{b}{2}(1-x^2)^2$ has a minimum at $x= x_{\rm min} = 1$ and is plotted in Fig.~(\ref{subfig:Mixed_Quartic_Pot}) for various values of the parameter $b$. By increasing $b$, one may increase the depth $\Delta V = b/2$ of the potential.  For $x > x_{\rm min}$, the potential being attractive helps the Brownian particle reach the absorbing site ($x=0$), while for $x < x_{\rm min}$, a barrier resists approach towards $x=0$. Chemical reactions are often visualised as barrier crossing processes where the rate of reaction is the rate of first passage over the barrier. The current potential is motivated by such processes. We are interested how SR can influence such barrier crossing. 

\begin{figure}[ht!]
\centering
\includegraphics[width = .40\textwidth,height=0.18\textheight]{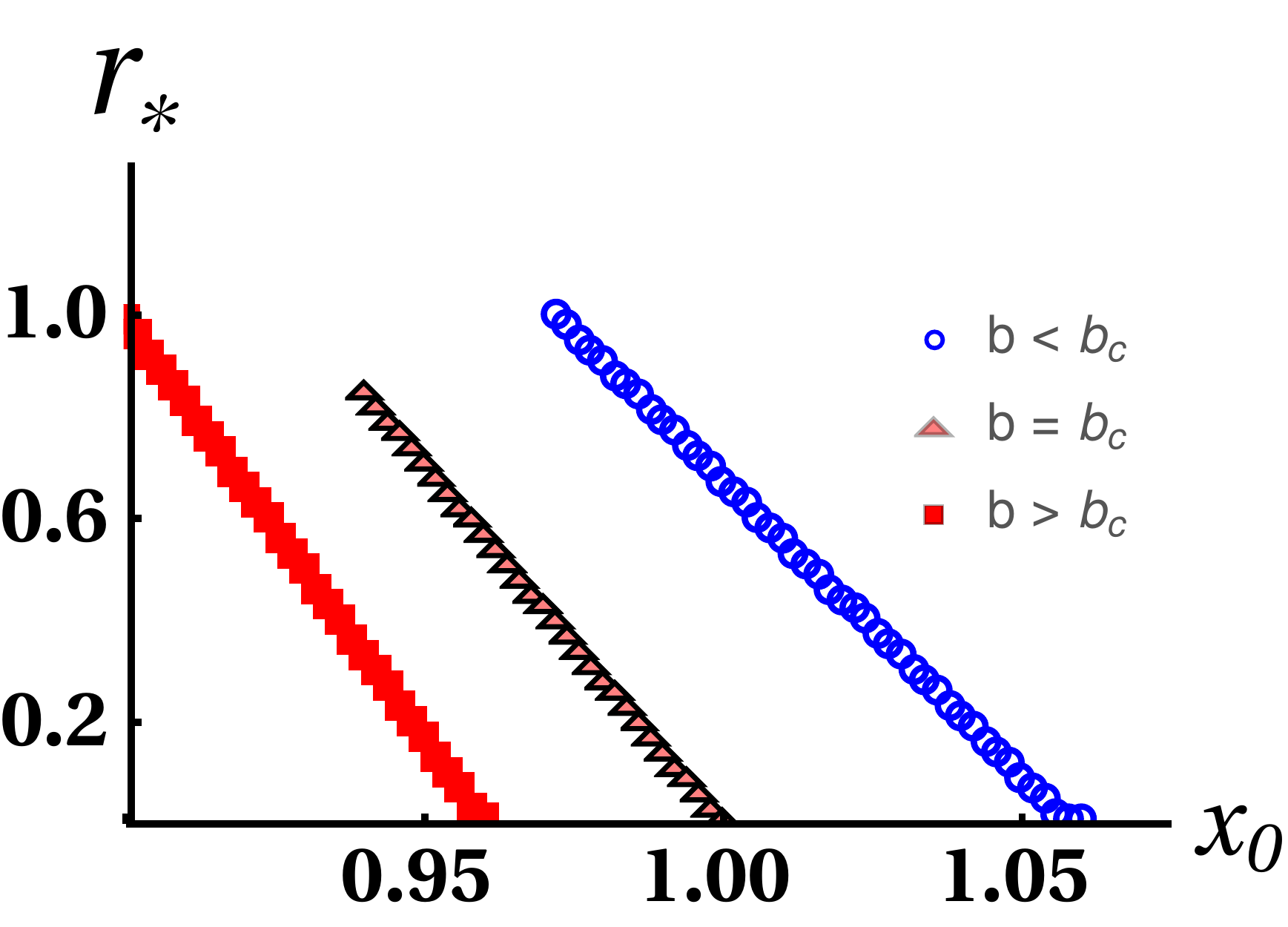}
\caption{We show $r_*$ vs. $x_0$ for different values of $b = 0.1696$, $b_c = 0.3392$, and $b = 0.5088$. The corresponding $x_{0c}$ values are $1.06$, $1.0$, and $0.96$, 
respectively. }
\label{fig:rx_Quartic_Mix_Pot}
\end{figure}

In this problem we vary the parameter $b$ so as to simultaneously increase the depth of the potential well, and also make it steeper for $x > x_{\rm min}$. For $b = 0$, to start with, we have a finite ORR ($r_*$) where $\langle T_r \rangle$ has a minimum, for an initial location and reset point $x_0 = x_{\rm min}$. The curiosity is to see that for the same $x_0$, with increasing $b > 0$, whether the ORR vanishes at some cutoff depth $b = b_c$, even when a barrier is present. Also for any fixed $b$, we study how the ORR vanishing behaves as a function of $x_0$.  
%%%%%%%%%%%%%%%%%%%%%%%%%%%%%%%%%%%%%%%%%%%%%%%%%%%%%%%%%%%%%%%%%%%%%%%%%%%%%%%%%%%%%%
%%%%%%%%%%%%%%%%%%%%%%%%%%%%%%%%%%%%%%%%%%%%%%%%%%%%%%%%%%%%%%%%%%%%%%%%%%%%%%%%%%%%%%

%%%%%%%%%%%%%%%%%%%%%%%%%%%%%%%%%%%%%%%%%%%%%%%%%%%%%%%%%%%%%%%%%%%%%%%%%%%%%%%%%%%%%%
%%%%%%%%%%%%%%%%%%%%%%%%%%%%%%%%%%%%%%%%%%%%%%%%%%%%%%%%%%%%%%%%%%%%%%%%%%%%%%%%%%%%%%
%%%%%%%%%%%%%%%%%%%%%%%%%%%%%%%%%%%%%%%%%%%%%%%%%%%%%%%%%%%%%%%%%%%%%%%%%%%%%%%%%%%%%%
%%%%%%%%%%%%%%%%%%%%%%%%%%%%%%%%%%%%%%%%%%%%%%%%%%%%%%%%%%%%%%%%%%%%%%%%%%%%%%%%%%%%%%
%\begin{figure}[hb!]
%\centering
%\includegraphics[width = .40\textwidth,height=0.20\textheight]{figures/Plot_Mixed_Quartic_Pot.pdf}
%\caption{The variation $V(x)$ with $x$ for various values of b.  The curves are for $b = 0.16959$, $b_c=0.33920$, and $b = 0.50879$.}
%\label{fig:Mixed_Quartic_Pot}
%\end{figure}
%%%%%%%%%%%%%%%%%%%%%%%%%%%%%%%%%%%%%%%%%%%%%%%%%%%%%%%%%%%%%%%%%%%%%%%%%%%%%%%%%%%%%%
%%%%%%%%%%%%%%%%%%%%%%%%%%%%%%%%%%%%%%%%%%%%%%%%%%%%%%%%%%%%%%%%%%%%%%%%%%%%%%%%%%%%%%

Using the numerical method discussed in this section, we find $\langle T_r \rangle$ for this problem with $x_0=1$ and $D=1$, and find that $r_*$ reduces with increasing  $b$ and vanishes for $b \geq b_c=0.3392\pm0.0001$.  Thus even in the presence of a resisting barrier, the rising steepness of potential (for $x > x_{\rm min}$) supersede its effect, and makes the advantage of resetting redundant beyond a point.  In Fig.~(\ref{subfig:Mixed_Pot_Tran}), we plot $r_*$ versus $(b_c-b)$ in log-log scale, and find a linear behaviour indicating the exponent $\beta =1$. For $b > b_c$, although there is no resetting advantage if reset point is $x_0 = 1$, if one has a reset to a nearer point to the absorbing site there is still advantage.  Similarly for $b < b_c$, the resetting advantage persists up to reset points $x_0 > 1$. These are quantitatively shown in Fig.~(\ref{fig:rx_Quartic_Mix_Pot}) -- we see that ORR vanishing happens at $x_{0c} < 1$ for $b > b_c$  and at $x_{0c} > 1$ for $b < b_c$.  

%%%%%%%%%%%%%%%%%%%%%%%%%%%%%%%%%%%%%%%%%%%%%%%%%%%%%%%%%%%%%%%%%%%%%%%%%%%%%%%%%%%%%%
%%%%%%%%%%%%%%%%%%%%%%%%%%%%%%%%%%%%%%%%%%%%%%%%%%%%%%%%%%%%%%%%%%%%%%%%%%%%%%%%%%%%%%
%\begin{figure}[hb!]
%\centering
%\includegraphics[width = .40\textwidth,height=0.20\textheight]{figures/Z2_K_Mixed_Quartic_Pot.pdf}
%\caption{Log-Log graph for $r_*$ vs. $b_c-b$. Here $D=1$ and $x_0=1$. A line with power $1$ is in good agreement with the data. }
%\label{fig:Mixed_Pot_Tran}
%\end{figure}
%

%%%%%%%%%%%%%%%%%%%%%%%%%%%%%%%%%%%%%%%%%%%%%%%%%%%%%%%%%%%%%%%%%%%%%%%%%%%%%%%%%%%%%%

\section{Discussion}

%A variety of first-passage processes in nature are guided by external fields. Bacterial cells move around with alternate phases of \emph{run} and \emph{tumble} \cite{berg_book}. In the presence of a chemical gradient, caused by nutrient sources, their motion gets biased towards the source \cite{koshland_pnas72}. Another instance is the capture of chromosomes by the spindle microtubules during mitotic cell-division. Experiments have revealed that chemical gradient around the chromosome often biases the microtubules to reach their target \cite{clarke_zhang_08}. Such examples serve as a motivation for studying stochastic processes under bias. In this paper we have studied prototype problems in which such bias competes with events of stochastic resetting. Our focus was to explore when can the advantage of resetting be annulled by the growing strength of the bias.

A variety of first-passage processes in nature are guided by external fields. Examples include bacterial cells performing \emph{run} and \emph{tumble} in presence of a spatial gradient of nutrients \cite{koshland_pnas_1972}, or, the movement of the spindle microtubules towards the chromosomes guided by a bio-chemical gradient \cite{clarke_zhang_nature_2008}. Such examples serve as a motivation for studying stochastic processes under bias. On the other hand SR is known to serve as an optimal strategy for first capture processes. In this paper we have studied few toy examples in which external bias competes with events of SR. We have obtained conditions under which the the advantage achieved by SR is annulled by the growing strength of the bias.

Here we have studied it systematically as a function of varying strength of a potential, that competes with SR at a constant rate. For a sufficiently strong potential, i.e. $k \geq k_c$, 
%, characterised by a dimensionless strength $K \propto k^{\frac{1}{n}}$, 
the ORR  vanishes. We derived the condition of transition, dependent on the moments of first passage time without resetting
%: $ \sigma^2  = \langle T \rangle^2 + 2 \langle T_{\rm on} \rangle \langle T \rangle $ 
(Eq.~\eqref{eq:ORR_condition}). Thus as the potential grows stronger and drives the particle more efficiently towards the capture site at origin, the fluctuations in first passage time characterised by $\sigma^2$ decreases until it matches the square of MFPT $\langle T \rangle^2$. Beyond that point, SR ceases to be of any extra assistance to the capture process. For processes with reset
followed by a stochastic time overhead, we show that the condition of transition is generalised to Eq.~\eqref{eq:ORR_condition_on}. Thus there is a limit to the advantage in first capture through SR, which  is set by the degree to which a system is biased towards capture  by an external force. 

Related to the general results discussed above, we have several specific observations. The non-dimensional critical potential strength $K_c$ varies monotonically with $n$ and reaches a constant value as $n \rightarrow \infty$ (Table~\ref{tab:Kc_behaviour}, section~\ref{sec5}). Also we observed that in the presence of finite refractory period ($\langle T_{\rm on} \rangle > 0$), the values of $K_c$ reduce
in comparison to the cases with zero refractory period (section IV).  The exponent associated with the power law form of vanishing $r_*$ appears quite universally to be $\beta  = 1$, owing to the ubiquitous analytic dependence of $r_*$ on $K$.  We derive explicit analytic forms, both in the absence  and presence  of refractory period, for $n=1$ and $n=\infty$. For $n=2$ our analysis is mostly analytical. For $n=3$ and $4$ we obtain the results by numerically solving the relevant differential equations to a high degree of accuracy.  The numerical method is applicable to any $n$ an in fact to any arbitrary potential whose first derivative exists. As an example, we studied a non-monotonic potential with  a barrier near the origin. We find that as a function of the depth of the potential, the ORR vanishes at and above a critical depth and the associated exponent $\beta = 1$. 

We believe that the transition studied in this paper and its mathematical criteria would be of general interest.   The numerical method that 
we developed could be useful in studying other similar problems related to first capture processes.    

{\it Acknowledgement}:  We are thankful to Sanjib Sabhapandit for useful discussions. AN acknowledges 
Science and Engineering Research Board (SERB), India (Project No. ECR/2016/001967) for financial support.

\appendix  
\section{ \label{APP:1} Detailed derivation of Eq.~(\ref{eq:MFPT_taylor}) }
Starting from Eq.~(\ref{eq:mean_Tr}) if we Taylor expand function $\tilde{F}(r)$ about point $r=0$, we have 
%\be
%\langle T_r \rangle = \frac{\big[1-\big(\tilde{F}(0)+r\frac{\partial \tilde{F}}{\partial r} |_{r=0}+\frac{r^2}{2!}\frac{\partial^2 \tilde{F}}{\partial r^2} |_{r=0}+\ldots\big)\big]}{r \big[\tilde{F}(0)+r\frac{\partial \tilde{F}}{\partial r} |_{r=0}+\frac{r^2}{2!}\frac{\partial^2 \tilde{F}}{\partial r^2} |_{r=0}+\ldots\big]},
%\label{eq:mean_Tr_expan}
%\ee
\begin{widetext}
\be
%\Scale[0.98]{
\langle T_r \rangle = \frac{1}{r}\bigg[\frac{1-\big(\tilde{F}(0)+r\frac{\partial \tilde{F}}{\partial r} |_{r=0}+\frac{r^2}{2!}\frac{\partial^2 \tilde{F}}{\partial r^2} |_{r=0}+\frac{r^3}{3!}\frac{\partial^3 \tilde{F}}{\partial r^3} |_{r=0}+ O(r^4)\ldots\big)}{\tilde{F}(0)+r\frac{\partial \tilde{F}}{\partial r} |_{r=0}+\frac{r^2}{2!}\frac{\partial^2 \tilde{F}}{\partial r^2} |_{r=0}+\frac{r^3}{3!}\frac{\partial^3 \tilde{F}}{\partial r^3} |_{r=0}+ O(r^4)\ldots}\bigg],
%},
\label{eq:App_mean_Tr_expan}
\ee
\end{widetext}
Here  $\tilde{F}(0)=1$, $\langle T \rangle = -\frac{\partial {\tilde{F}}}{\partial r} |_{r=0}$, $\langle T^2 \rangle = \frac{\partial^2 {\tilde{F}}}{\partial r^2} |_{r=0}$, $\langle T^3 \rangle = -\frac{\partial^3 {\tilde{F}}}{\partial r^3} |_{r=0}$ and so on. Hence in terms of the different moments we 
may rewrite the MFPT as 
\be
\langle T_r \rangle =  \bigg[\frac{\langle T \rangle-\frac{r}{2!}\langle T^2 \rangle+\frac{r^2}{3!}\langle T^3 \rangle-\frac{r^3}{4!}\langle T^4 \rangle+O(r^4)\ldots}{1-r\langle T \rangle+\frac{r^2}{2!}\langle T^2 \rangle-\frac{r^3}{3!}\langle T^3 \rangle+ O(r^4)\ldots}\bigg].
\label{eq:App_mean_Tr_expan1}
\ee
Binomial expansion of the denominator in Eq.~\eqref{eq:App_mean_Tr_expan1} leads to 
\be
\Scale[0.9]{
\begin{aligned}
\langle T_r \rangle &= \langle T \rangle  - r \bigg(\frac{\langle T^2 \rangle}{2} - \langle T \rangle^2\bigg) + r^2 \bigg(\frac{1}{6} \langle T^3 \rangle +\langle T \rangle^3-\langle T \rangle\langle T^2 \rangle\bigg) \\ 
&+~r^3\bigg(-\frac{\langle T^4 \rangle}{4!} +\frac{\langle T^3 \rangle\langle T \rangle}{3}+\frac{\langle T^2 \rangle^2}{4}-\frac{3\langle T^2 \rangle\langle T \rangle}{2}+\langle T \rangle^4\bigg) \\
&+~O(r^4)\ldots
\label{eq:App_MFPT_taylor1}
\end{aligned}
}
\ee
To obtain the  optimal resetting rate $r_*$,  we take a derivative of Eq.~\eqref{eq:App_MFPT_taylor1} and 
set $d \langle T_r \rangle/dr |_{r=r_*} = 0$. This leads to a quadratic equation for $r_*$: 
\be
\begin{aligned}
 cr^2_*+2br_*+a=0,
\label{eq:App_MFPT_taylor2}
\end{aligned}
\ee
where $a$, $b$ and $c$ are given by 
\be
\Scale[0.8]{
\begin{aligned}
a &= -\bigg(\frac{\sigma^2-\langle T \rangle^2}{2}\bigg).\\
b &= \bigg(\frac{1}{6} \langle T^3 \rangle -\sigma^2\langle T \rangle\bigg); \\
c &= 3\bigg(-\frac{\langle T^4 \rangle}{4!} +\frac{\langle T^3 \rangle\langle T \rangle}{3}+\frac{\langle T^2 \rangle^2}{4}-\frac{3\langle T^2 \rangle\langle T \rangle}{2}+\langle T \rangle^4\bigg).\\
\end{aligned}
}
\ee
The solution of Eq.~\eqref{eq:App_MFPT_taylor2} is $ r_* = \dfrac{-b\pm\sqrt{b^2-ac}}{c} $. 
In the immediate neighbourhood of the transition point, as $a\to 0$, the value of $r_*$ may be approximated as: 
\begin{align}
 r_* &\approx\qquad {} \frac{-b}{c}\pm\frac{b}{c}\bigg(1-\frac{ac}{2b^2}\bigg) \nonumber \\
      &=\qquad {} -\frac{a}{2b}, \qquad {}\text{for the positive root.}
\end{align}
The above equation is same as Eq.~\eqref{eq:r_*} in the main text. 
Furthermore, Eq.~\eqref{eq:ORR_condition} for ORR transition point  is obtained by setting $a=0$.  
%\appendix*

\bibliography{apssamp_v4}% Produces the bibliography via BibTeX.
%%%%%%%%%%%%%%%%%%%%%%%%%%%%%%%%%%%%%%%%%%%%%%%%%%%%%%%%%%%%%%%%%%%%%%%
%%%%%%%%%%%%%%%%%%%%%%%%%%%%%%%%%%%%%%%%%%%%%%%%%%%%%%%%%%%%%%%%%%%%%%%
%%%%%%%%%%%%%%%%%%%%%%%%%%%%%%%%%%%%%%%%%%%%%%%%%%%%%%%%%%%%%%%%%%%%%%%
%%%%%%%%%%%%%%%%%%%%%%%%%%%%%%%%%%%%%%%%%%%%%%%%%%%%%%%%%%%%%%%%%%%%%%%
%%%%%%%%%%%%%%%%%%%%%%%%%%%%%%%%%%%%%%%%%%%%%%%%%%%%%%%%%%%%%%%%%%%%%%%
%%%%%%%%%%%%%%%%%%%%%%%%%%%%%%%%%%%%%%%%%%%%%%%%%%%%%%%%%%%%%%%%%%%%%%%
%%%%%%%%%%%%%%%%%%%%%%%%%%%%%%%%%%%%%%%%%%%%%%%%%%%%%%%%%%%%%%%%%%%%%%%
%%%%%%%%%%%%%%%%%%%%%%%%%%%%%%%%%%%%%%%%%%%%%%%%%%%%%%%%%%%%%%%%%%%%%%%
%%%%%%%%%%%%%%%%%%%%%%%%%%
%%%%%%%%%%%%%%%%%%%%%%%%%%
\end{document}